\documentclass[pdflatex,sn-mathphys]{sn-jnl}

\usepackage{verbatim}
\usepackage{subfigure}

\jyear{2021}%

\theoremstyle{thmstyleone}%
\theoremstyle{thmstyletwo}%

\theoremstyle{thmstylethree}%
\usepackage{pifont}

\newcommand{\CO}{CO$_2$}
\definecolor{rcolor}{rgb}{0.85,0.37,0.25}
\definecolor{mscolor}{rgb}{0.12,0.25,0.75}
\newcommand{\rb}[1]{{}} 
\definecolor{berndcolor}{rgb}{0.00,0.75,0.25}
\newcommand{\bernd}[1]{{}} 
\newcommand{\ms}[1]{{}} 

\newcommand{\etal}[0]{~et\,al.}

\definecolor{qcolor}{rgb}{0.89, 0.029, 0.83}
\newcommand{\qn}[1]{{}} 

\definecolor{scolor}{rgb}{0.09, 0.029, 0.83}
\newcommand{\sff}[1]{{}} 

\definecolor{gcolor}{rgb}{0.09, 0.029, 0.83}
\newcommand\gr[1]{{}} 

\newcommand{\boxb}[0]{Box\,B}
\newcommand{\boxa}[0]{Box\,A}
\newcommand{\boxc}[0]{Box\,C}

\newcommand{\tff}[0]{transfer function}
\newcommand{\Tff}[0]{Transfer function}
\newcommand{\spatiotemp}[0]{spatio-temporal}

\newcommand{\rqs}[1]{$\mathbf{(Q_#1)}$}
\newcommand{\tsks}[1]{$\mathbf{(T_#1)}$}

\definecolor{austin}{HTML}{1f77b4}
\definecolor{csiro}{HTML}{ff7f0e}
\definecolor{delft-DARSim}{HTML}{2ca02c}
\definecolor{delft-DARTS}{HTML}{d62728}
\definecolor{heriot-watt}{HTML}{9467bd}
\definecolor{lanl}{HTML}{8c564b}
\definecolor{melbourne}{HTML}{e377c2}
\definecolor{stanford}{HTML}{7f7f7f}
\definecolor{stuttgart}{HTML}{bcbd22}
\definecolor{experimentrun1}{HTML}{17becf}
\definecolor{experimentrun2}{HTML}{000075}
\definecolor{experimentrun3}{HTML}{ffe119}
\definecolor{experimentrun4}{HTML}{bfef45}

\newcommand{\austin}[1]{\textcolor{austin}{#1}}
\newcommand{\csiro}[1]{\textcolor{csiro}{#1}}
\newcommand{\delftDARSim}[1]{\textcolor{delft-DARSim}{#1}}
\newcommand{\delftDARTS}[1]{\textcolor{delft-DARTS}{#1}}
\newcommand{\heriotwatt}[1]{\textcolor{heriot-watt}{#1}}
\newcommand{\lanl}[1]{\textcolor{lanl}{#1}}
\newcommand{\melbourne}[1]{\textcolor{melbourne}{#1}}
\newcommand{\stanford}[1]{\textcolor{stanford}{#1}}
\newcommand{\stuttgart}[1]{\textcolor{stuttgart}{#1}}
\newcommand{\experimentrunA}[1]{\textcolor{experimentrun1}{#1}}

\begin{document}

\title[Visual Analysis for Fluid Flow in Porous Media Simulation Ensemble]{
Visual Ensemble Analysis of Fluid Flow in Porous Media across Simulation Codes and Experiment
}

\author*[1]{\fnm{Ruben} \sur{Bauer}}\email{ruben.bauer@visus.uni-stuttgart.de}

\author[1]{\fnm{Quynh Quang} \sur{Ngo}}\email{quynh.ngo@visus.uni-stuttgart.de}

\author[1]{\fnm{Guido} \sur{Reina}}\email{guido.reina@visus.uni-stuttgart.de}

\author[3]{\fnm{Steffen} \sur{Frey}}\email{s.d.frey@rug.nl}

\author[2]{\fnm{Bernd} \sur{Flemisch}}\email{bernd.flemisch@iws.uni-stuttgart.de}

\author[4]{\fnm{Helwig} \sur{Hauser} \email{helwig.hauser@uib.no}}

\author[1]{\fnm{Thomas} \sur{Ertl}}\email{thomas.ertl@vis.uni-stuttgart.de}

\author[1]{\fnm{Michael} \sur{Sedlmair}}\email{michael.sedlmair@visus.uni-stuttgart.de}

\affil*[1]{\orgdiv{VISUS}, \orgname{University of Stuttgart}, \orgaddress{
\country{Germany}}}

\affil[2]{\orgdiv{IWS}, 
\orgname{University of Stuttgart}, 
\orgaddress{
\country{Germany}}}

\affil[3]{
\orgname{University of Groningen}, \orgaddress{
\country{the Netherlands}}}

\affil[4]{
\orgname{University of Bergen}, \orgaddress{
\country{Norway}}}

\abstract{
We study the question of how visual analysis can support the  comparison of \spatiotemp{} ensemble data of liquid and gas flow in porous media. 
To this end, we focus on a case study, in which nine different research groups concurrently simulated the process of injecting \CO{} into the subsurface. 
We explore different data aggregation and interactive visualization approaches to compare and analyze these nine simulations. 
In terms of data aggregation, one key component is the choice of similarity metrics that define the relation between the different simulations. 
We test different metrics and find that a fine-tuned machine-learning based metric provides the best visualization results. 
Based on that, we propose different visualization methods. 
For overviewing the data, we use dimensionality reduction methods that allow us to plot and compare the different simulations in a scatterplot. 
To show details about the \spatiotemp{} data of each individual simulation, we employ a space-time cube volume rendering. 
We use the resulting interactive, multi-view visual analysis tool to explore the nine simulations and also to compare them to data from experimental setups. 
Our main findings include new insights 
into ranking of simulation results with respect to experimental data,
and the development of gravity fingers in simulations.
}

\keywords{
Porous media, fluid flow, visual analytics, benchmark study, simulation ensemble
}

\maketitle

\section{Introduction}
Injecting \CO{} 
into subsurface reservoirs might be a key approach in the future to mitigate climate change~\cite{Bachu:2007:CSC,Pacala:2004:SWS,Metz:2005:ISR}.
Toward this approach, however, it is fundamental to gain a better understanding of fluid flow and transport in porous media,
an area which has attracted substantial attention in many research fields~\cite{Bear:2018:MPF,Sahimi:2011:FTP,Kamrava2021}. Concerning geological carbon storage, large experimental efforts have been undertaken at potential storage sites to collect information about, for example, the geology and formation fluids as well as their respective dynamics~\cite{Lindeberg:2009:DCS,Niemi:2016:HES}. 
These efforts are costly though and often limited to very specific conditions. 
To overcome these problems, simulation studies such as \cite{Class:2009:BSP} have become  popular, taking advantage of the increasing computational capabilities. 
To validate the respective simulation models, they should be compared to existing experimental data.
In the absence of such ground truth experiment data, however, the validation necessitates the careful exploration and analysis of simulations with different settings to capture all potential phenomenal patterns of fluid flow in porous media. 

The main goal of our work is to explore how interactive visualization can support exploring, comparing, and analyzing different simulations in this domain. 
To this end, we focus on a benchmark study of geological storage of \CO{} in the subsurface~\cite{Nordbotten2022BenchmarkStudy,Flemisch:2023:SAS}. 
In a larger project consortium, nine different research groups around the globe were tasked with simulating this process. 
The result of each individual simulation is \spatiotemp{} data (2D+time) that predicts the behavior of \CO{} flow starting from a joint, pre-defined condition. 
More precisely, the output of each simulation constitutes 2D spectral images containing saturation and concentration of \CO{} in each cell of a uniform Cartesian grid discretizing the 2D spatial domain, recorded in ten-minute intervals. 
This \spatiotemp{} data is complemented by additional measurables such as the pressure at a specified location or the \CO{} mass integrated over a specific region, each providing a separate scalar time series.

After the individual simulations were run, the main challenge is now to explore the resulting ensemble of simulations, to compare similarities and differences between them, and to set them into context to the underlying experimental simulation that was conducted along with the simulations.
This set of exploratory tasks leans itself toward a visual analysis approach in general~\cite{munzner2014visualization}, and ensemble visualization in particular~\cite{Sedlmair2014VisualParamSpaceAnalysis,DBLP:journals/cgf/FofonovL19}. 
In an interdisciplinary team of visualization experts and a porous media domain expert, we set out to better understand how ensemble visualization can benefit this area, and how respective visualizations should be designed. 
Over the course of 1.5 years, we followed a joint design study process~\cite{sedlmair2012design} and explored different data aggregation and visualization approaches for the data at hand.

The main idea behind most ensemble visualizations is to derive a quantitative similarity metric that allows to relate different ensemble members to each other and to visually compare them in the same space~\cite{survey2019}. 
The choice of metrics largely depends on the domain application though, and so far no universal similarity metrics exist for fluid flow data in porous media. 
We found, for instance, that  metrics such as the Euclidean and Wasserstein distance, which are commonly used in porous media simulation, become increasingly inaccurate with increasing time steps and number of dimensions, and might fail to capture important details in the data. 
To address this issue, we leverage a machine-learning based approach to define the similarity between ensemble members, which we adapt for the domain problem at hand. 
Still, there seems to be no one-size-fits-all solution. 
Different metrics have different benefits and drawbacks. 
Here, we can again leverage interactive data visualization that allows to see and analyze the data through these different angles.  

The similarity metrics can then be used in an overview visualization. 
To this end, we split simulation results into different timeslots (patches) and project them as time curves~\cite{bach2015time} into a 2D scatterplot. 
This visualization allows us to find similarities between different simulations at different times. 
We additionally embed the experimental data into the same space, which allows us to put the nine different simulations into a global context. 
We extend this overview visualization with different detail visualizations. 
We use a space-time cube volume rendering~\cite{Bach2017DescFrameworkForSpaceTimeCubes} to present the full \spatiotemp{} simulation results of each ensemble member.
Another juxtaposed view is used to display the respective scalar time series, and interaction allows to dive deeper into specific questions. 

When using the resulting interactive visualization tool on the benchmark study data, we discovered several novel findings that suggest further investigations for domain scientists, such as comparisons of the length, shape and development behavior of gravity fingers and interesting quantifications of similarities between simulation results and experimental data.

In summary, we make the following contributions:
\begin{enumerate}
    \item We propose a visual comparative analysis approach utilizing a variety of similarity metrics for ensemble data of simulating fluid flow in porous media.
    \item Using our approach, we explore data from a benchmark study, revealing new insights about the underlying domain.
\end{enumerate}

\section{Background and Related Work} 
This section provides background on the simulation benchmark study and related work.
We first provide some background on the modelled \CO{} injection process, and briefly summarize the data generation process of the benchmark study.
We then discuss various examples of how related work has dealt with the visual analysis of similar simulation ensembles.

\subsection{Benchmark Study}
\label{subsection:benchmark_study_and_porous_media}
We focus on analyzing simulation ensemble data provided from a recent benchmark study of Nordbotten\etal~\cite{Nordbotten2022BenchmarkStudy}.
The benchmark study is inspired by a climate change research problem, which concerns the injection of \CO{} into subsurface reservoirs. 
Subsurface reservoirs are geological structures below ground that are suitable for long-term storage of fluids.
They usually consist of layers with different porosity and permeability such as a coarse-grained highly permeable region which is suited for storing a large amount of fluid, and a fine-grained low permeable caprock above which prevents the stored fluid from escaping.
In these naturally occurring subsurface reservoirs, possibly large amounts of \CO{} can be injected and captured first below the caprock.
Over time, more and more of the \CO{} dissolves into the formation water and the \CO{}-saturated water sinks downward, increasing long-term storage security~\cite{Bachu:2007:CSC,Metz:2005:ISR,Pacala:2004:SWS}.
This process of convective mixing is driven by the density difference of the original formation water and the \CO{}-saturated water and usually manifests itself in the form of so-called ``gravity fingers''~\cite{Nordbotten2022BenchmarkStudy}.
 
For such large-scale real-world scenarios, it is important to assess potential risks and opportunities by trying to model and forecast them~\cite{Pruess:2004:CIB,Class:2009:BSP}.
Even with a good understanding of the complex physical processes during and after injection of \CO{} into porous media, the lack of knowledge about the precise conditions in the subsurface introduces many uncertainties.
As experiments are costly and field-scale real-time measurements are prohibitive due to the targeted long time spans, this problem is often addressed by uncertainty quantification approaches which require running many forward simulations that cover different conditions~\cite{Walter:2012:BMR,Sun:2019:DAU}.

The main goal of the benchmark study was to ``provide a full-physics validation of the state-of-the-art simulation capabilities''~\cite{Nordbotten2022BenchmarkStudy} for such \CO{} injection processes.
With the help of an experimental rig for repeated multiphase 2D flow experiments\footnote{The FluidFlower Concept: Operating Flow Rigs \url{https://fluidflower.w.uib.no/large-scale/}}, a laboratory-scale \CO{} injection experiment was conducted which served as reference for the benchmark study. 
In the beginning of the study, the most important boundary conditions, like geometry, operation process of the injection, and other physical parameters of the experimental rig, were specified and provided to nine simulation expert groups.
These groups were then asked to model, simulate, and forecast the actual experiments that were performed with the rig, but without having access to the experimental data.
The experiments were run by an independent experimental group and serve as ground truth for the analysis and validation of the simulation data.

Appropriate analysis and comparison methods are required that allow to inspect and compare different simulations with each other and with the experimental data.
This is a non-trivial problem, especially for \spatiotemp{} and multivariate data.
We address this problem with our visual approach which is designed to support the analysis of such ensemble data. 

\subsection{Related Work}\label{subsection:related_work}
From the data types and design methodology perspective, we review in this section related work about visual comparative analysis of ensemble \spatiotemp{} data. 
Particularly, we focus on similarity metrics and visual analysis methods. 

\subsubsection{Similarity Metrics for Ensemble Spatiotemporal Data}
Visual parameter space/ensemble analysis~\cite{Sedlmair2014VisualParamSpaceAnalysis} normally concerns how parameter configurations influence the outcome of a simulation system by comparing ensemble members. 
One of the main challenges in \spatiotemp{} ensemble analysis is finding a suitable similarity/distance metric for an assisted or automated ensemble member comparison.
While a manually performed visual comparison of \spatiotemp{} data can be employed for pattern recognition and comparison of complex time series in the details, it is not suitable for large data sets such as simulation ensembles.
A similarity metric could support providing an abstraction overview of the ensemble in form of a scatterplot. 
The overview is the context that enables us to efficiently and automatically compare, filter, rank, and cluster ensemble members before performing a time-intensive manual analysis and comparison of-and-between individual members.
In our visual approach, we utilize various similarity metrics, including an unsupervised machine learning-based approach.

Tkachev\etal~\cite{Tkachev2022S4}~presented S4---an ML-driven similarity metric, which is based on the assumption that spatial proximity implies similar behavior. 
We utilize their approach for our use-case to learn a similarity metric on the provided simulation data. 
However, we change the patch sizes and network size to address the problem of overfitting with respect to our limited amount of available data.
In a recent work, Huesmann and Linsen proposed SimilarityNet~\cite{Huesmann2022SimilarityNet}, which is a ML model trained in an autoencoder-like fashion on a generated phantom dataset to learn to encode arbitrary \spatiotemp{} data into a 1D space representation that preserves \spatiotemp{} behavior.
Due to the limitation of only producing 1D embeddings, we exclude this approach from our design, as we use 2D overviews for other metrics. 

\subsubsection{Visual Analysis for Ensemble Spatiotemporal Data}
We refer here to a survey by Obermaier and Joy~\cite{ObermaierFutureChallenges2014} which categorized existing ensemble visualizations into ``feature-based visualization'', and ``location-based visualization''.
Our visual approach supports analysis components that fall into both categories.
A recent survey of Wang\etal~\cite{WangSurvey2019} categorizes ensemble visualization from two perspectives: the proposed visualization techniques, and the involved analytic tasks.
Technique-wise, we consider multivariate data and mainly address linked juxtaposed views including two views for direct volume rendering.
Regarding tasks, we address most of the mentioned tasks directly, except ``clustering" and ``parameter analysis".

Existing work also provides different design applications for a variety of ensemble \spatiotemp{} data types and respective domain application tasks. 
Demir\etal~~\cite{DemirComp3dEnsVis2020} presented a chart-based approach showing statistical properties of 3D volume ensemble members to support comparative analysis. 
H{\"o}llt\etal~\cite{Hoellt2016VisualAnalysisofReservoirSimEns} contributed a visual analysis tool for reservoir simulation ensembles, utilizing statistical measures.  
Potter\etal~\cite{PotterEnsembleVis2009} built a visualization framework focusing on statistical measures of simulation ensemble data. 
Bach\etal~\cite{Bach2017DescFrameworkForSpaceTimeCubes} provided a descriptive framework for temporal data visualizations based on generalized space-time cubes. 
Fofonov and Linsen~\cite{Fofonov2018MultiVisAVA} focused their work on multi-run physical simulation data, analyzing the impact of initial conditions and parameter settings on simulation results. 
Our ensemble consists of only few members with a variety of different parameter settings which makes it unsuitable for a quantitative parameter space analysis.
We focus on an interactive visual comparison analysis of the ensemble members using variety of similarity metrics instead of statistical properties of ensemble data.

\section{Problem Characterization}
We characterize our problem by providing the description of the data, 
our research questions, and the respective analysis tasks. 

\subsection{Data}\label{subsection:data}
We consider an ensemble of nine different simulations of \CO{} injection processes into porous structures from nine research groups that participated in the benchmark study, labeled \textcolor{austin}{\textit{austin}}, \textcolor{csiro}{\textit{csiro}}, \textcolor{delft-DARSim}{\textit{delft-DARSim}}, \textcolor{delft-DARTS}{\textit{delft-DARTS}}, \textcolor{lanl}{\textit{lanl}}, \textcolor{heriot-watt}{\textit{heriot-watt}}, \textcolor{melbourne}{\textit{melbourne}}, \textcolor{stanford}{\textit{stanford}}, and \textcolor{stuttgart}{\textit{stuttgart}}.
From each of the reported simulation data, we consider a series of 2D spatial maps, which represent the recorded \CO{} saturation and concentration values for the first 24 hours in ten-minute intervals.
A saturation value is the ratio of the volume occupied by the gaseous phase to the available void space for each considered reporting cell, while a concentration value indicates the mass of dissolved \CO{} per volume of liquid phase in that cell.

\begin{figure}[hbt]
\centering
\includegraphics[width=0.95\textwidth]{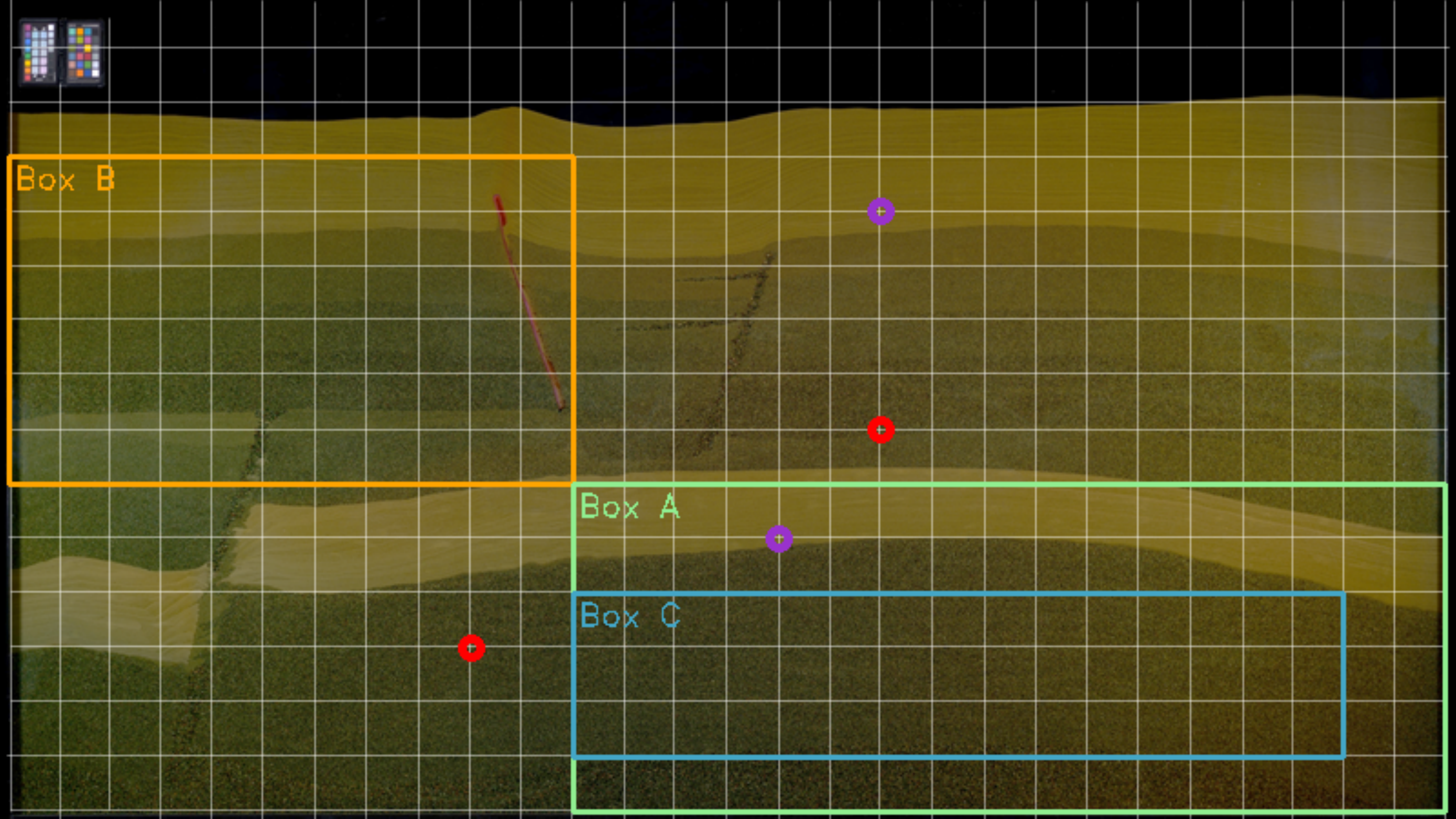}
\caption{Photograph image of the benchmark geometry with overlaid laser grid \cite[Figure 8]{Nordbotten2022BenchmarkStudy}. The red circles indicate the injection points, while the purple circles depict the pressure sensors. Boxes A-C correspond to regions for the evaluation of different system response quantities.}
\label{fig:benchmark_geometry}
\end{figure}

At each time step, additional measurables such as local pressure values from sensors and integrated quantities for three different regions of interest are reported, which we consider as time series information.
The three regions of interest correspond to predefined rectangular regions in the benchmark study: \boxa{}, \boxb{}, and \boxc{}, see \autoref{fig:benchmark_geometry}.
They are defined to capture and express specific features and events during the simulations and experiments. 
In particular, the research groups were requested to provide the total
mass of \CO{} inside the domain, pressure at two locations, phase composition
in Boxes A and B, as well as convection in Box C. 

Besides the simulation groups, there was also one experimental group which performed the actual reference experiments with the setup mentioned in~\autoref{subsection:benchmark_study_and_porous_media}.
We have access to the segmentation data of four experiment runs which we integrated in our visual approach.
In contrast to the simulation data, which provides saturation and concentration values at each grid cell, the experimental data only provides ternary information about whether there is (i) only pure water, (ii) water with dissolved \CO{} or (iii) also gaseous \CO{} in a cell, due to the difficult process of post-processing the experimental data.
In the post-processing, saturation and concentration values have to be derived from photographs of the experiment by analyzing the \CO{}-induced coloring of the water.
The time series data was not available to us for the experimental runs.

\subsection{Research Questions and Tasks}
\label{sub:tasks}
We work alongside a domain expert who has been working with flow and transport processes in porous media for $15$ years with whom we had a regular bi-weekly meeting.  
In a series of multiple focus group meetings, we jointly derived a set of research questions that should be possible to analyze with the target visualization tool.

\begin{itemize}
    \item $\mathbf{Q_{1}}$: 
    How similar are simulation outcomes across research groups and what are the differences? %
    \item $\mathbf{Q_2}$: Which groups' simulation outcomes match most closely the available experimental data? %
    \item $\mathbf{Q_3}$: Is there any correlation between the scalar time series data and the dynamics of the spatial saturation and concentration distribution? %
    \item $\mathbf{Q_{4}}$: Concerning features of particular interest: When do the ``fingers''~(\ref{subsection:benchmark_study_and_porous_media}) reach a certain length? %
    When is the spilling point reached? (Here, the spilling point is the location where \CO{} ``spills'' over the edge of the modelled reservoir~(\autoref{fig:benchmark_geometry}, \boxa{}) after reaching its maximum capacity of mobile free \CO{} gas.)
\end{itemize}
We derive a list of tasks for the analysis of fluid flow on an ensemble of porous media simulations.

\noindent
\textbf{Comparative analysis of ensemble members:} %
$\mathbf{(T_1})$-compare simulations of different groups on a certain level of abstraction, to answer \textbf{Q1}. %
$\mathbf{(T_2)}$-rank simulations with regard to the experimental results, to answer \textbf{Q2}. %
$\mathbf{(T_3)}$-find correlations between saturation and concentration over time, to answer \textbf{Q3}. %
\textbf{Spatial detail tasks:}
$\mathbf{(T_4)}$-find areas having an interesting \CO{} concentration/saturation.
\textbf{Temporal detail tasks:}
$\mathbf{(T_5)}$-find out when certain events happen. 
Both $\mathbf{T_4}$ and $\mathbf{T_5}$ are to address \textbf{Q4}. 
\section{Visual Analysis Approach}\label{section:methodology}

\begin{figure*}
    \centering
    \includegraphics[width=\textwidth]{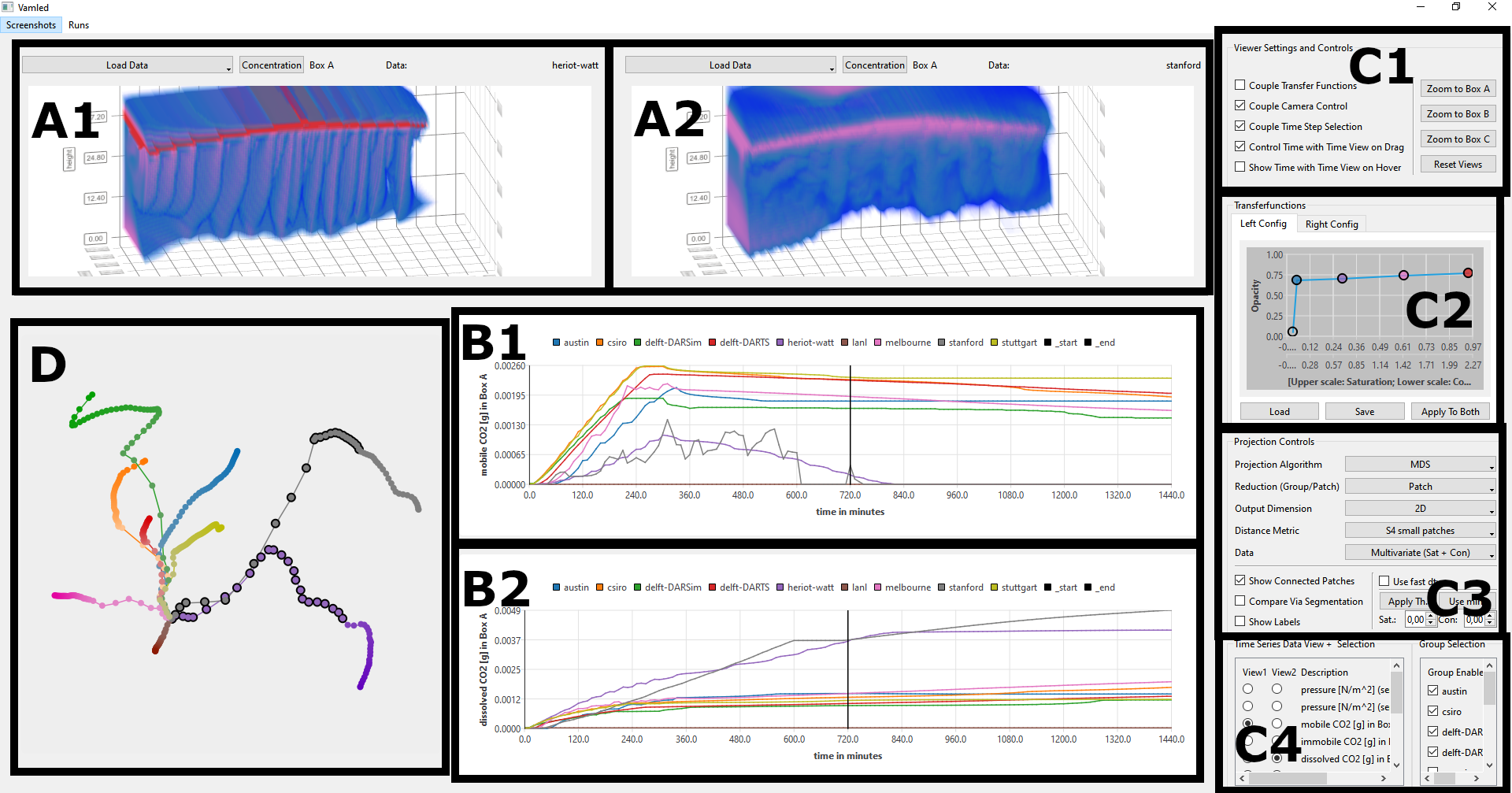}
    \caption{Overview of our user interface. (\textbf{A1}\ \&\ \textbf{A2}): Space-time cube renderings with dedicated selected group and visualized variable. (\textbf{B1}\ \&\ \textbf{B2}): Time series plots of additional measurables. (\textbf{C1}): Viewer settings and interaction controls. (\textbf{C2}): Transfer function to map saturation and concentration to color and opacity. (\textbf{C3}): Projection controls for view (\textbf{D}). (\textbf{D}): Projection of the ensemble based on algorithm, metric, and data selected in (\textbf{C3}).}
    \label{fig:teaser}
\end{figure*}

In our design process, we considered the requirements and derived tasks from~\autoref{sub:tasks} as well as the provided data as described in~\autoref{subsection:data}.
As such, our visual analysis approach will follow the overview first, details on demand mantra~\cite{Schneiderman1996Mantra}. 
Below, we first introduce the data processing employed for our visual approach. 
Second, we discuss the similarity metrics that we use to compare simulation outcomes. 
Third, we describe the proposed \texttt{data representation}, \texttt{interaction}, and \texttt{controls} components of our visual analysis tool. 
Fourth, we finally provide an example workflow for a visual analysis using our visual approach. 

\subsection{Data Processing}\label{subsection:data-pipeline}
The data processing happens at two stages: the pre-processing required for our visual approach, and the extraction of so-called \emph{patches}. 

\subsubsection{Pre-Processing}\label{subsubsection:data-preprocessing}
To make it easier to observe \spatiotemp{} patterns, we propose to use a static view of the data from each group, namely space-time cube visualization~\cite{hagerstrand_what_1970}. 
The spatial maps then have to be densely packed and transformed to a volume (a 3D texture) in which the uniform grid of the spatial maps as well as their time components map to indices of the resulting volume.

For a proper visual comparison, the resulting volume should be of equal size and cover the same geometric and temporal extents for all ensemble members. 
However, the amount of time steps and the geometric extents of the spatial maps between different groups in the first $24$\,h varies slightly, though mostly by a difference of one time step or grid cell.
To align the data in the volumes, we first fill in missing time steps by repeating the data of the preceding time step, if available.
If there is no previous time step, we fill the time step with zeroes.
We set the target geometric extents of the volumes to the extents as described by the benchmark description.
The width and height of the resulting volumes represent the benchmark geometry from $x=0.005$\,m to $x=2.855$\,m, and from $y=0.005$\,m to $y=1.225$\,m with a step size of $0.01$m in $x$- and $y$ direction.
The depth of the volume represents the time from $t=0$\,s to $t=8640$\,s (=$24$\,h) with a step size of $600$\,s (=$10$\,min).

We perform the same procedure to fill missing values in the time series as we do for the spatial maps.
Missing time steps at the beginning become zero.
The remaining gaps are filled by repeating the previous existing value.
The resulting time series span the full range from $t=0$\,h to $t=24$\,h with one measurement every ten minutes.

\subsubsection{Processing of Patches}\label{subsubsection:datamodel}
We utilize the S4~\cite{Tkachev2022S4} ML model as a similarity metric for \spatiotemp{} behavior (see~\autoref{subsubsection:similarity-metrics}).
The S4 model trains on so-called \spatiotemp{} ``patches'' of the data.
A patch is a subset in the original \spatiotemp{} data. 
Considering the spatial maps of the benchmark study as a 3D volume with $x$, $y$, and $t$ dimension, a patch in this 3D volume context could be any 3D cuboid in it.
We empirically chose the temporal patch size (dimension $t$) to be always three, which equals $30$\,minutes in our data. 
This size allows the model to integrate the temporal component during the computation of latent space features without introducing too much change between two consecutive non-overlapping patches.
We utilize two versions of the S4 model with each a different spatial patch size, which we further specify in~\autoref{subsubsection:similarity-metrics}.

\subsection{Similarity Metrics}\label{subsubsection:similarity-metrics}
In the visualization community, similarity metrics are applied to evaluate how similar data items are. 
The similarity information is often fed to a dimensionality reduction (DR)~\cite{TSNE-vanDerMaaten2008, UMAP-McInnes2018, MDS-kruskal1964nonmetric} that provides a 2D mapping of the data items. 
This mapping's outcome is then represented in a scatterplot to serves as an overview of the dataset in a visual analysis pipeline. 

In this work, multiple similarity metrics are utilized to potentially capture more information with respect to some potential features of the simulation outcomes \tsks{1}.
Each metric computes the similarity between two patches with respect to their multivariate facets in general.
By using similarity metrics to automatically compute a similarity value between simulations of different groups, we can compare them in an abstract manner without inspecting the \spatiotemp{} manually, but to view it in a projection instead.
The projection then provides hints for a more detailed, manual analysis of individual patches and groups.

\paragraph*{Euclidean Distance and Manhattan Distance}
Euclidean distance and Manhattan distance are two of the most popular but simple metrics to compute distances between feature vectors.
In our case, to compute the distance between two patches, we linearize both patches to a feature vector and apply the corresponding metric.
While this is a common process, it is not exactly suited to use Euclidean distance or Manhattan distance to compare \spatiotemp{} data with patterns that might change in size, (\spatiotemp{}) position, or orientation, which physical phenomena often exhibit.
If two patches capture the exact same pattern, a small change in any of those properties may result in a large distance between these two patches, since the indices of the corresponding elements in the feature vector that capture the patterns might change completely.
However, we will show that Euclidean distance still yields reasonable results on the benchmark study ensemble~(\autoref{subsection:sim-metrics-cap-patterns}).

\paragraph*{ML Model for Comparing Spatiotemporal Behavior}
S4 is a ML model for ``Self-Supervised Learning of Spatiotemporal Similarity'', which was recently proposed by Tkachev et al.~\cite{Tkachev2022S4}. 
We expect the S4 model to be a more advanced and better-suited metric for comparing data by its \spatiotemp{} behavior.
The model has to be trained first, before it can be applied to the data.
It is trained on patches of the data and learns to encode them into a latent-space feature representation in which two vectors are close by Manhattan distance if their corresponding patches have similar \spatiotemp{} behavior.
The training exploits the assumption that two patches that are close in space and time are also close in terms of their \spatiotemp{} behavior and vice versa, and thus can be applied to unlabeled data.

We trained the model on a spatially downsampled volume by a factor of two.
The patch size is chosen to be the remaining full spatial size and a temporal size of three time steps.
During inference, we compute the non-overlapping patches of each group and use S4 to compute a $64$\,feature vector for each patch. %
The S4 distance between two patches is then the Manhattan distance between their corresponding feature vectors.
Even though we expect this model to be a more meaningful similarity metric for \spatiotemp{} data, our results will show that choosing the full spatial size as patch size yields too few data points and variations for the amount of trainable parameters which results in strong overfitting.
Thus, we trained another model with smaller patch size by spatially subdividing the patches, and reducing the amount of trainable parameters by slightly adjusting the models' hyperparameters to decrease the risk of strong overfitting.

To distinguish between both models in our analysis, we will refer to the first one as just ``S4'', and to the latter one as ``S4 with subdivided patches''.
For the ``S4 with subdivided patches'', we chose the spatial patch size to roughly capture the average finger width and length of the simulations in the first $24$\,h.
We hope that besides increasing the amount of data points and reducing the risk of overfitting, this allows the model to better capture local features, especially the finger development.
During inference with ``S4 with subdivided patches'', to now compare two patches of full spatial size with each other, we first spatially subdivide these patches in non-overlapping sub-patches and compute the sum of the distances between sub-patches at the same positions instead.

\paragraph*{Wasserstein Distance}
The Wasserstein distance~\cite{kantorovich1960WassersteinDistance} is well-known for flow simulation data.
It is a measure of the distance between two probability distributions, which in our case, is the distribution of concentration or saturation values in a single patch.
First, we compute a histogram of a patch's saturation values and a histogram of a patch's concentration values by binning the values into $256$ bins from min to max saturation/concentration.
We divide the histograms by the total amount of elements in the patch, which yields a probability distribution of the saturation and concentration values in a patch.
Now we can apply the Wasserstein distance between the two saturation distributions and concentration distributions.
We average the two distances to get a combined final result of our metric.
We emphasize that this Wasserstein distance is different from the one employed in \cite{Flemisch:2023:SAS}, where two-dimensional distributions over the spatial domain and the corresponding Euclidean metric are used. 

\subsection{Data Views}\label{subsection:views-and-controls}
In this subsection, we describe the different views representing our proposed data visual encodings and respective interaction controls in showing how they relate to each other as well as to our defined tasks in~\autoref{sub:tasks}.
An overview of our implementation instance for all views is provided in~\autoref{fig:teaser}.
Besides the similarity plot, which provides an overview of the similarity of the full ensemble, 
we provide views to link the projected ensemble members or their patches to the actual data from spatial maps 
or time series information. 

\subsubsection{Similarity View}\label{subsubsection:similarity-view}
Following the overview first, zoom-and-filter, details-on-demand mantra~\cite{Schneiderman1996Mantra}, we project the ensemble's similarity in a similarity plot to provide an overview of the full ensemble.
The similarity metric information among patches or ensemble members (by getting average similarity metrics of sub-patches in the ensemble members) is fed to a DR technique to derive the similarity plot in forms of a 2D scatterplot. 
Each point in the scatterplot represents either one patch or one ensemble member (in the case of using average similarity metrics of sub-patches in ensemble members), and the distances among the points visually reflect the similarity among the patches or ensemble members. 
The similarity plot in \autoref{fig:teaser}\,(\textbf{D}) provides hints as to which groups are outliers or if small clusters have formed, decreasing the mental workload of manually going through the whole ensemble to compare them~\tsks{1}.

Furthermore, inspired by Machicao et al.~\cite{MACHICAO20211}, we include the experimental data in the projection, which allows us to visually rank the simulation groups by comparing their similarities to the experimental data~\tsks{2}.
As experimental data is only available in the form of segmentation maps, we have to transform the simulation data into segmentation maps to get them into the same format for comparison. 
The segmentation maps of the experimental groups were computed by choosing a certain threshold in the image analysis when analysing the original photographs of the experiment.
This threshold relates to how much concentration or saturation exists in the image grid cell.
Therefore, we also choose a threshold to transform the simulation data into segmentation maps before computing the similarity matrix with our selected similarity metric.
We set the default threshold to consider a grid cell to contain \CO{} saturation or \CO{} concentration to close to zero ($0.001$) to avoid the segmentation of actually empty cells due to numerical errors.

Like similarity metrics, DR algorithms may differ in the revealed features and quality of results~\cite{TSNE-vanDerMaaten2008, UMAP-McInnes2018, MDS-kruskal1964nonmetric}.
We chose to integrate three popular DR algorithms that are often used for projecting data into 2D space.
We figure multidimensional scaling (MDS)~\cite{MDS-kruskal1964nonmetric} to be one of the most important DR algorithms in our context, as it tries to preserve distances between data points.
Besides MDS, we also integrate uniform manifold approximation and projection (UMAP)~\cite{UMAP-McInnes2018} and
t-distributed stochastic neighbor embedding (t-SNE)~\cite{TSNE-vanDerMaaten2008}. 
These two DR techniques try to preserve the neighborhoods of data points. 

As mentioned above, we provide two types of similarity plots in this work. 
Choosing ``group'' mode (in the case of using average similarity metrics of sub-patches in ensemble members) aggregates the data to project only one point per group, which makes it easier to detect clusters or outliers in the ensemble.
The ``patch'' mode projects all patches and gives hints about the temporal similarity between the groups.
\autoref{fig:similaritymetricgroups} shows MDS projections of the ensemble using the ``group'' option, while~\autoref{fig:similaritymetricpatchs} shows MDS projections of the ensemble using the ``patch'' option.

\subsubsection{Space-Time Cube View}\label{subsubsection:space-time-cube}
\begin{figure*}
    \subfigure[\Tff{} that shows low concentration.]{\includegraphics[width=0.40\textwidth]{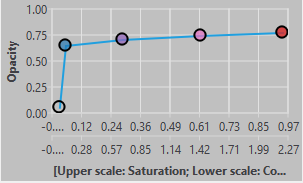}}
    \subfigure[\stanford{Stanford} \boxa{} with \tff{} (a).]{\includegraphics[width=0.56\textwidth]{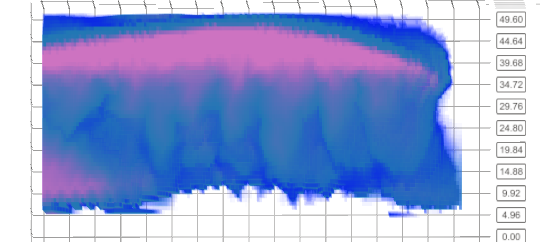}}
    \subfigure[\Tff{} without low concentration.]{\includegraphics[width=0.40\textwidth]{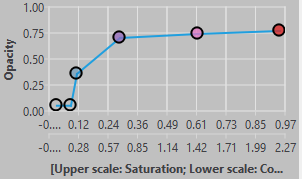}}
    \hspace{0.3cm}
    \subfigure[\stanford{Stanford} \boxa{} with \tff{} (d).]{\includegraphics[width=0.56\textwidth]{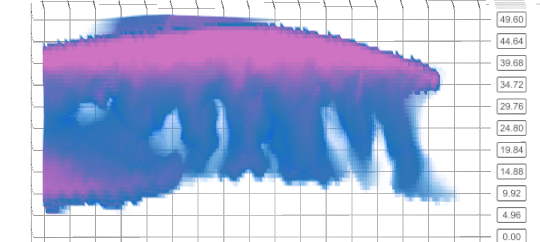}}
    \caption{
    \stanford{Stanford} spatial maps renderings for \boxa{} with two different \tff{}s.
    By selecting a \tff{} which omits low concentration values, the shape of the higher concentrated fingers becomes visible.
    }
    \label{fig:transferfunction-filter-values}
\end{figure*}
To support focusing on details for \spatiotemp{} patterns of each ensemble member, we propose to use a static view representing a series of 2D maps as a volume with time as the third dimension. 
The space-time cube rendering~\cite{Bach2017DescFrameworkForSpaceTimeCubes} renders the processed volume (\autoref{subsubsection:data-preprocessing}) of either saturation or concentration information of data. %
The saturation or concentration values of the space-time cube are mapped to colors with respective luminance and opacity defined by a color transfer function, see~\autoref{fig:teaser}\,(\textbf{A1}\,\&\,\textbf{A2}). %
We argue that this representation of the data provides better insight than juxtaposing 2D spatial data representation images in a sequence as it smoothly maintains the temporal evolution pattern of the data. 

To make it more domain-application friendly, we propose to allow the user to interactively change/define the transfer function while analyzing the data using the view. 
Particularly, the user can choose colors and opacity for any saturation or concentration values in our data by using the \tff{} diagram, see~\autoref{fig:teaser}\,(\textbf{C2}).
Values that do not have an explicit value have the linearly interpolated color and opacity between their left and right point applied to them.
By properly configuring the \tff{}, the user can highlight or hide ranges of saturation and concentration values of the space-time cube volume.
In~\autoref{fig:transferfunction-filter-values}, we show two examples of a space-time cube rendering with different \tff{}s that each show different concentration values.
Besides the interactive transfer function, the view also provides fundamental interaction such as rotation, zoom-in, and zoom-out to make it easier for the user to analyze and study the data cube. 

The space-time cube view alone can support the user to target tasks \tsks{4} and \tsks{5}. 
Being static, the view supports comparing the spatial distribution of concentration/saturation of different time steps globally~\tsks{4}. 
From the comparison, the user can look for the area of interest and can determine the time steps at which certain events happen~\tsks{5}. 

We propose to use two space-time cube views in our design to target \tsks{1} and \tsks{3}, see~\autoref{fig:teaser}\,(\textbf{A1}\,\&\,\textbf{A2}). 
When the two views are used to represent concentration and saturation information of the same ensemble data, the user can analyze and study if there exists a correlation/relationship between the two types of information~\tsks{3}. 
Together with the similarity view, if the two views are used to represent one type of information for two ensemble members, the user can comparatively analyze and study in detail the two ensemble members~\tsks{1}. 

\subsubsection{Line Charts View}\label{subsubsection:time-series-data}
We use line charts to visualize the time series data of the given three regions of interest, i.e., regions \boxa{}, \boxb{}, and \boxc{}.
Each line chart is used to represent one measurable of one group over time.
The line charts with different colors for different groups are superimposed in one view, see~\autoref{fig:teaser}\,(\textbf{B1}\,\&\,\textbf{B2}). 
This allows us to visually compare the development of a measurable among different groups over time. 

We propose to contain two juxtaposed views in which each can visualize one of those line charts for one measurable at a time, see~\autoref{fig:teaser}\,(\textbf{B1}\,\&\,\textbf{B2}).
This provides a convenient side-by-side comparison pattern over simulation groups of two different measurables of different regions.
As a decluttering mean, we provide the user the ability to interactively select and deselect one or more groups and to choose the selected measurable of interest that are shown in the views, see~\autoref{fig:teaser}\,(\textbf{C1}).
To target~\tsks{3}, we propose to link the saturation and concentration of the spatial maps that are visualized in the space-time cubes to the corresponding time series measurables mobile\,\CO{} and dissolved\,\CO{}. 
The detail of the interaction operation will be presented in~\autoref{subsubsection:interaction}. 
By analyzing line charts of different regions of interest, the user can also identify events of interest and respective regions related to the provided measurables, i.e., targeting \tsks{4} and \tsks{5}.

\subsubsection{Interaction}
\label{subsubsection:interaction}
\ms{a video would be really nice to illustrate the tool and the interactions.}
Besides the interaction operations that we designed for each aforementioned view, we use brushing and linking to coordinate among views. 
The similarity view is linked to the space-time cube view and the time series view to support the top-down analysis approach~\tsks{1}.
The user can select two patches or two ensemble members of interest to be displayed in the two space-time cube views, e.g.,~\autoref{fig:teaser}\,(\textbf{A1}\,\&\,\textbf{A2}), to compare them in detail. 
When the similarity displays the patches, hovering over the patches' representations will slice the space-time cube of the corresponding group to the selected patches.
The similarity view is further linked to the time series view and vice versa.
Hovering over a patch's representation in the similarity view will select the range of this patch in the line charts view, e.g.,~\autoref{fig:teaser}\,(\textbf{B1}\,\&\,\textbf{B2}).
Selecting a time range in the time series views will highlight the corresponding patches of the groups in the similarity view which are also currently visible in the space-time cube view.

By providing a convenient interaction to inspect the measurables and spatial maps at different time steps simultaneously, the interlinking of the three views effectively targets tasks \tsks{3}, \tsks{4}, and \tsks{5}. 
Linking between the time-cube view and the line charts view allows the user to analyze correlation between time series input and saturation and concentration information~\tsks{3}. 
Meanwhile, linking the similarity view with the other two views allows the user to observe the overview pattern before focusing on details about some specific group, the specific time step, and the specific spatial region that constitutes the pattern, i.e., targeting \tsks{4}, and \tsks{5}. 
\subsection{Workflow}\label{subsection:workflow}
Based on the overview first -- zoom and filter -- details on demand mantra, we propose the following workflow with our visual analysis approach. 

\subsubsection*{Step 1: Analyze the Similarity View}

\noindent 
Regarding overview first, the user can take a look at the similarity view showing the ``group'' mode of the ensemble. 
The similarity plot encodes the different groups by color.
The user can identify clusters of some groups being closer to each other than others, or identify outliers.
If the user is interested in the temporal development of the ensemble members or how they diverge throughout the simulation, the ``patch" mode can be enabled. 
Hovering over a patch of one group can highlight the corresponding patches of the other group at the same time step (Figure~\ref{subfig:2dmdsS4smallEnhanced}). %
By doing that, the user can see how fast and when two groups diverge. %

If the user is interested in how well simulation data compares to the experimental results, the experimental data can be included in the projection results.

The user can further inspect the overview under the changing similarity metrics and DR techniques to see how the results change.
%
\subsubsection*{Step 2: Link Similarity View To Detail Views}

\noindent 
From observing the similarity pattern of ensemble members in the similarity view with ``group" mode, the user can compare the details pair of the members using space-time cube views.  
The space-time cube rendering shows the whole outer shape of the simulation at once and makes it easy to roughly estimate if the two simulations behave visually similar over time.
Inspecting the space-time cube might verify if one outlier in the projection indeed has a completely different behaviors in the spatial maps than the others.

If the similarity view is in ``patch" mode, the user can navigate the resulting projection via zooming to inspect early time steps that greatly overlap.
By hovering over the patches, the user can compare the corresponding actual spatial maps in the space-time cube views.
The time series view reveals the corresponding time steps of the hovered patches, allowing the user to see whether close patches are also similar in the given time series data.
\subsubsection*{Step 3: Analyze Each Ensemble Member in Detail Views}

\noindent
Having got a broad overview of the ensemble, the user can analyze the ensemble members in more detail.
The user can investigate the difference between concentration and saturation data by comparing the respective two time-space cube views~(Figures~\ref{subfig:boxbsatdelft-darsim},~\ref{subfig:boxbcondelft-darsim}). 
Next, the user can inspect how the \CO{} concentration evolves over time in the simulations by interactively defining the transfer function.
When the user is interested in the \CO{} concentration and wants to better inspect the finger development in the different boxes of the simulation geometry, they can zoom to the respective boxes by spatially slicing the volume to contain only the box of interest.
After that the user can select the line chart that shows the corresponding dissolved\,\CO{} in our box of interest.
The user can select the range where the line chart first indicates existing dissolved\,\CO{} in the box of interest up to when the amount of dissolved\,\CO{} does not seem to change anymore.
This slices the space-time cube temporally to the selected range.
With that, the user can now inspect the space-time cube and its early stages of finger development in the box of interest.

\section{Results}\label{section:results}
In this section, we provide anecdotes showing the answers to the aforementioned research questions in \autoref{sub:tasks} as well as other findings by using our visual analysis approach. 
We first provide the results related to comparing the groups and finding similarities/differences by using our similarity metrics \rqs{1}. 
The outcomes related to ranking with respect to experimental data \rqs{2} is mentioned after that. 
Finally, we present several findings related to \rqs{3} and \rqs{4}. 

\subsection{Different Similarity Metrics Capture Different Patterns of the Dataset -- \textbf{Q1}}\label{subsection:sim-metrics-cap-patterns}
\begin{figure*}
    \subfigure[S4\label{subfig:s4nexp}]{\includegraphics[width=0.19\textwidth]{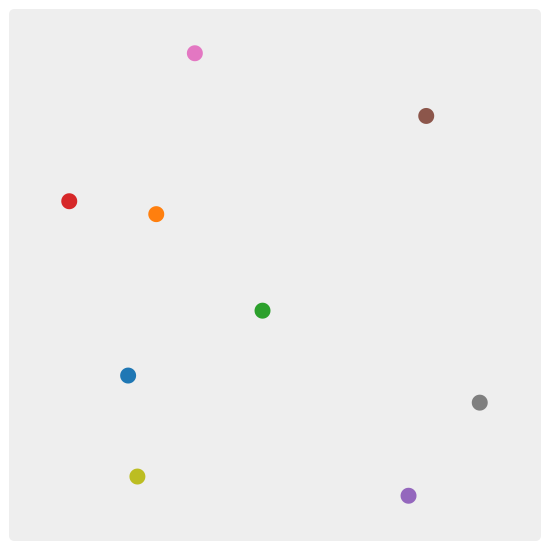}}
    \subfigure[S4 w. subdiv. patches\label{subfig:s4smallnexp}]{\includegraphics[width=0.19\textwidth]{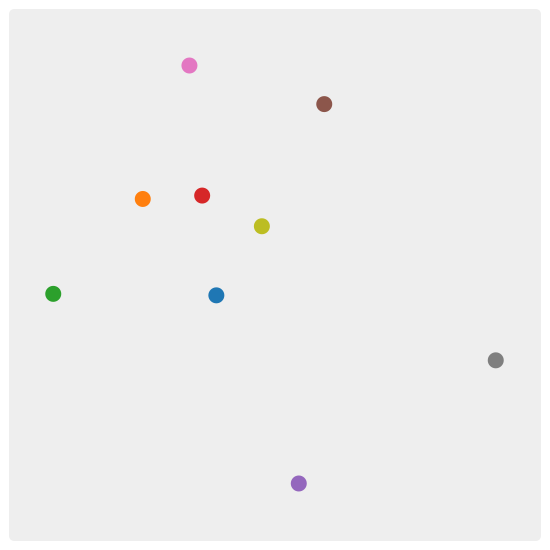}}
    \subfigure[Euclidean\label{subfig:ednexp}]{\includegraphics[width=0.19\textwidth]{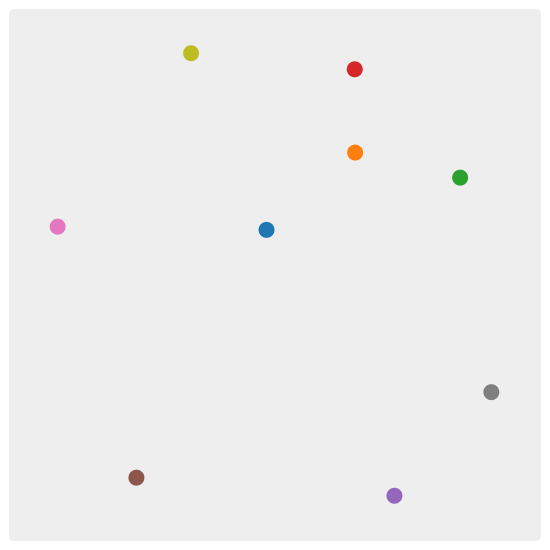}}
    \subfigure[Manhattan\label{subfig:mannexp}]{\includegraphics[width=0.19\textwidth]{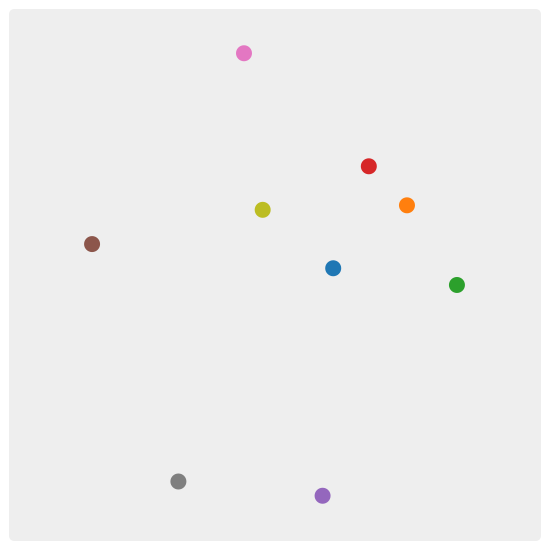}}
    \subfigure[Wasserstein\label{subfig:wssnexp}]{\includegraphics[width=0.19\textwidth]{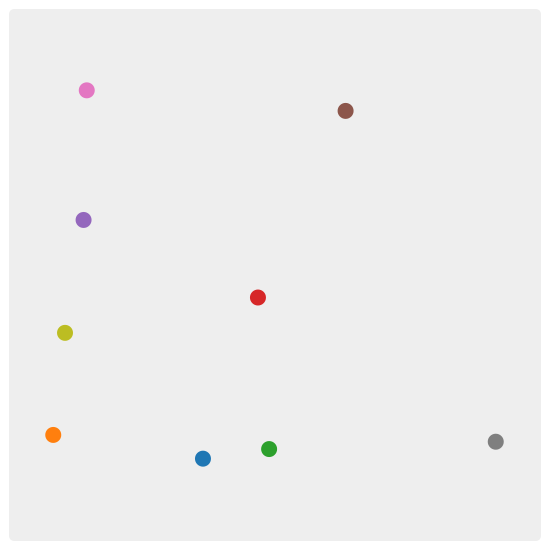}}
    \subfigure[S4 w. exp.\label{subfig:s4wexp}]{\includegraphics[width=0.19\textwidth]{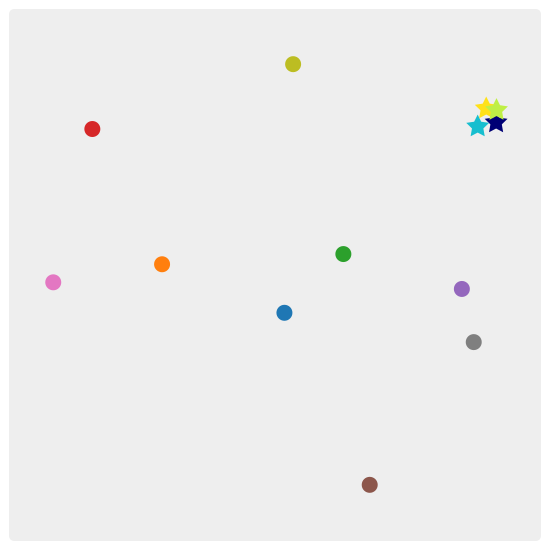}}
    \subfigure[S4 w. subdiv. patches w. exp.\label{subfig:s4smallwexp}]{\includegraphics[width=0.19\textwidth]{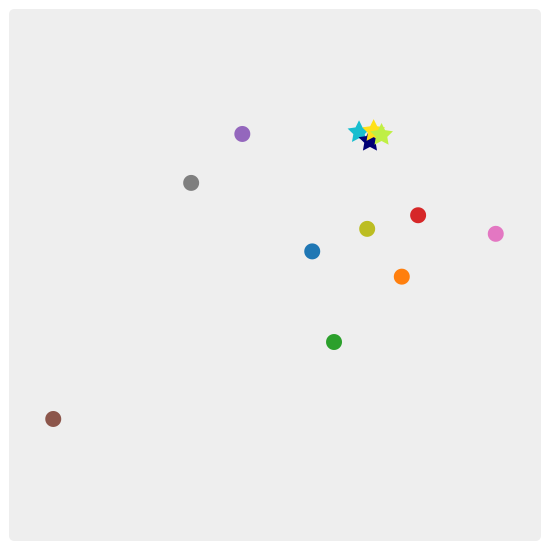}}
    \subfigure[Euclidean w. exp.\label{subfig:edwexp}]{\includegraphics[width=0.19\textwidth]{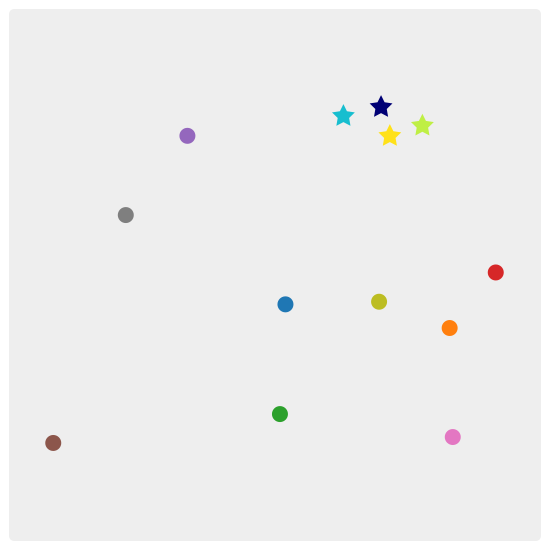}}
    \subfigure[Manhattan w. exp.\label{subfig:manwexp}]{\includegraphics[width=0.19\textwidth]{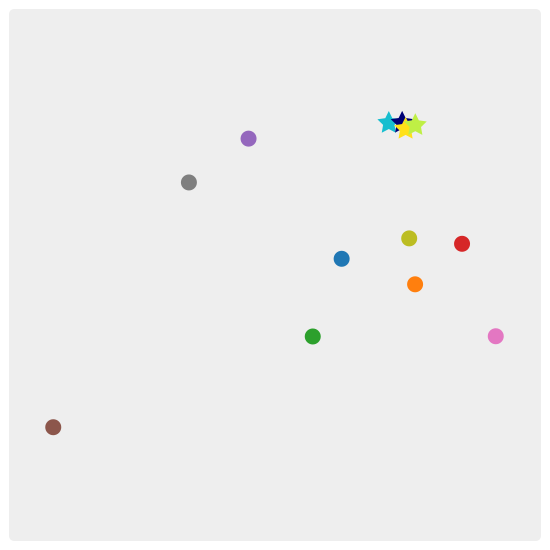}}
    \subfigure[Wasserstein w. exp.\label{subfig:wsswexp}]{\includegraphics[width=0.19\textwidth]{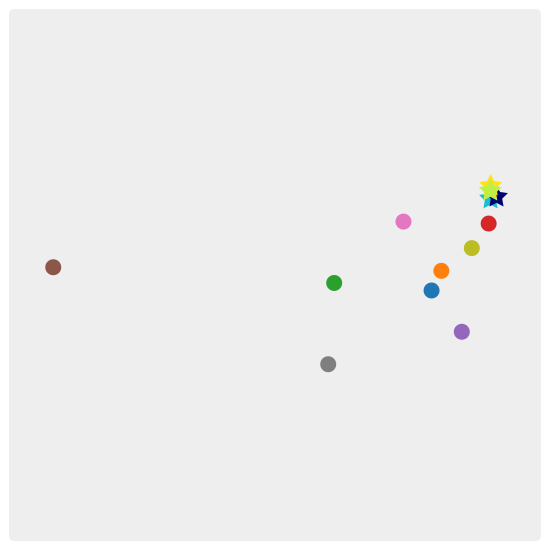}}
    \begin{tabular}{r@{ }l r@{ }l r@{ }l r@{ }l r@{ }l}
        \textcolor{austin}{\textbullet} & austin & \textcolor{csiro}{\textbullet} & csiro & \textcolor{delft-DARSim}{\textbullet} & delft-DARSim & \textcolor{delft-DARTS}{\textbullet} & delft-DARTS & \textcolor{lanl}{\textbullet} & lanl \\
        \textcolor{heriot-watt}{\textbullet} & heriot-watt & \textcolor{melbourne}{\textbullet} & melbourne & \textcolor{stanford}{\textbullet} & stanford & \textcolor{stuttgart}{\textbullet} & stuttgart \\
        \textcolor{experimentrun1}{\ding{72}} & exp. run1 & \textcolor{experimentrun2}{\ding{72}} & exp. run2 & \textcolor{experimentrun3}{\ding{72}} & exp. run3 & \textcolor{experimentrun4}{\ding{72}} & exp. run4
    \end{tabular}
    \caption{
    (a)-(e): 2D plots of the ensemble simulation spatial maps for different similarity metrics with aggregated patches (``group" mode).
    (f)-(j): same as (a)-(e) but using the segmented spatial maps and including the experimental data in the projection.
    The experimental data is plotted as star shapes instead of circles.
    The segmentation threshold is $0.001$.
    }
    \label{fig:similaritymetricgroups}
\end{figure*}

\begin{figure*}
\begin{minipage}[c]{0.34\textwidth}
    \subfigure[2D projection via MDS and S4 with subdivided patches.\label{subfig:2dmdsS4small}]{\includegraphics[width=1\textwidth]{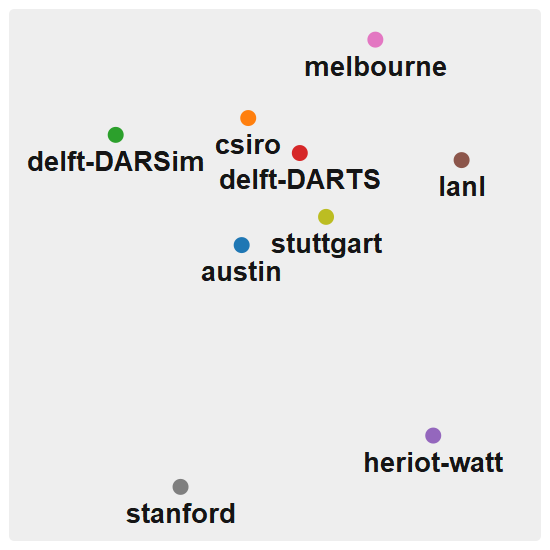}}
\end{minipage}
\begin{minipage}[c]{0.66\textwidth}
    \subfigure[\austin{austin}]{\includegraphics[width=0.49\textwidth]{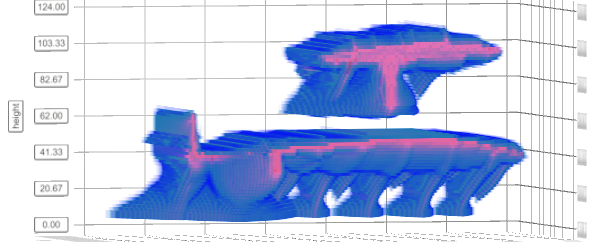}}
    \subfigure[\stuttgart{stuttgart}]{\includegraphics[width=0.49\textwidth]{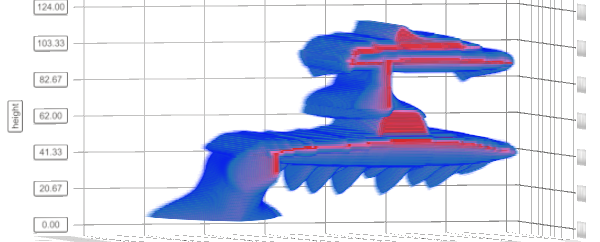}}
    \subfigure[\heriotwatt{heriot-watt}]{\includegraphics[width=0.49\textwidth]{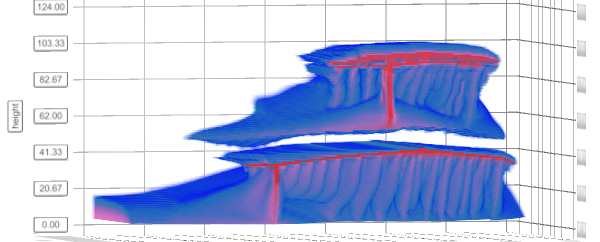}}
    \subfigure[\stanford{stanford}]{\includegraphics[width=0.49\textwidth]{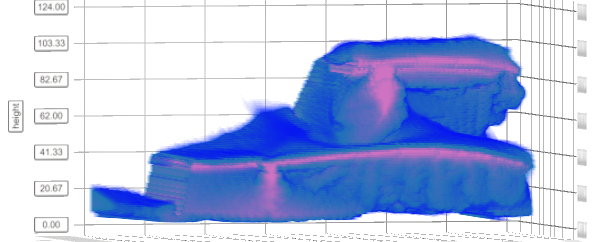}}
\end{minipage}
    \caption{
    The visually perceivable differences between the space-time cube renderings (b)-(e) aligns with the computed and projected distances of the spatial maps in (a).}
    \label{subsubsection:similaritfig:similarity_compared_to_spatial_mapsy-view}
\end{figure*}
\begin{figure*}
    \subfigure[S4\label{subfig:patch-s4nexp}]{\includegraphics[width=0.19\textwidth]{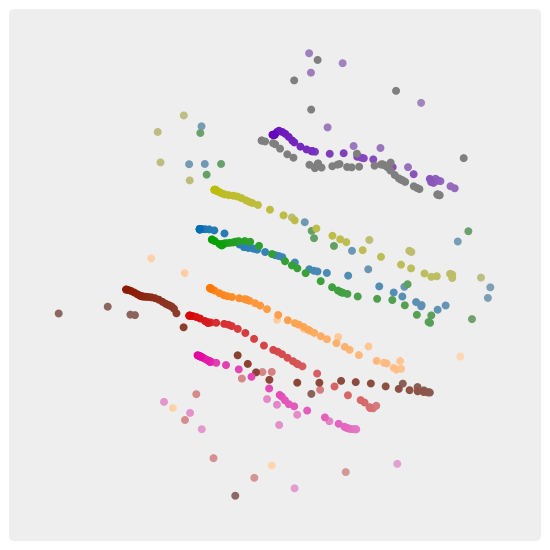}}
    \subfigure[S4 w. subdiv. patches\label{subfig:patch-s4smallnexp}]{\includegraphics[width=0.19\textwidth]{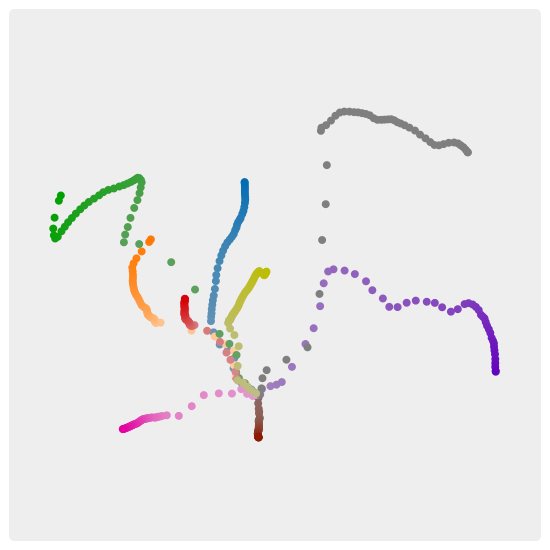}}
    \subfigure[Euclidean\label{subfig:patch-ednexp}]{\includegraphics[width=0.19\textwidth]{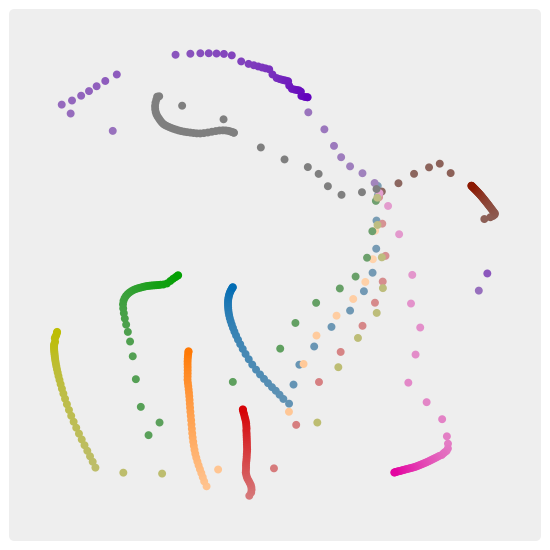}}
    \subfigure[Manhattan\label{subfig:patch-mannexp}]{\includegraphics[width=0.19\textwidth]{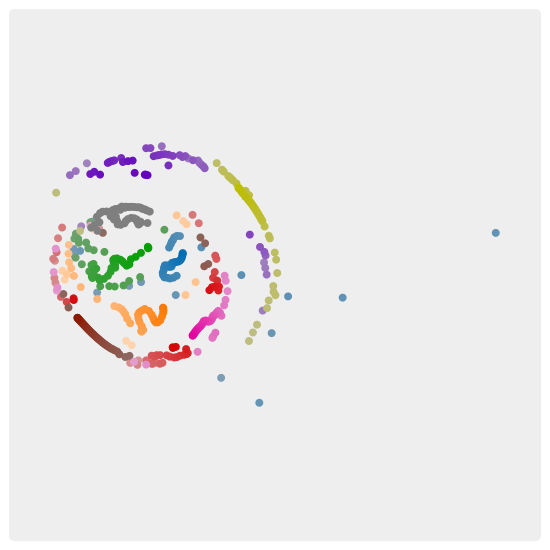}}
    \subfigure[Wasserstein\label{subfig:patch-wssnexp}]{\includegraphics[width=0.19\textwidth]{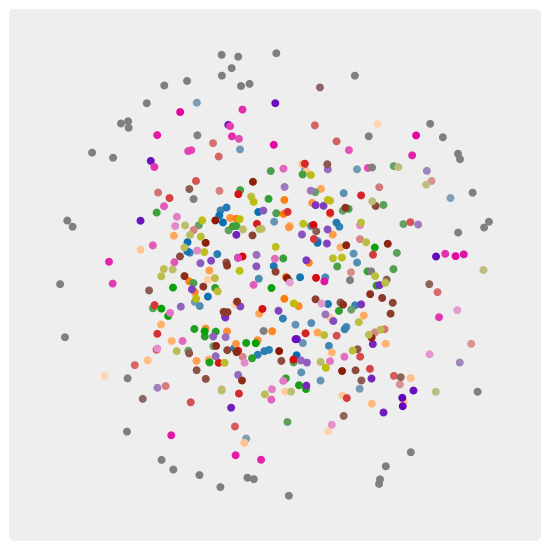}}
    \subfigure[S4 w. exp.\label{subfig:patch-s4wexp}]{\includegraphics[width=0.19\textwidth]{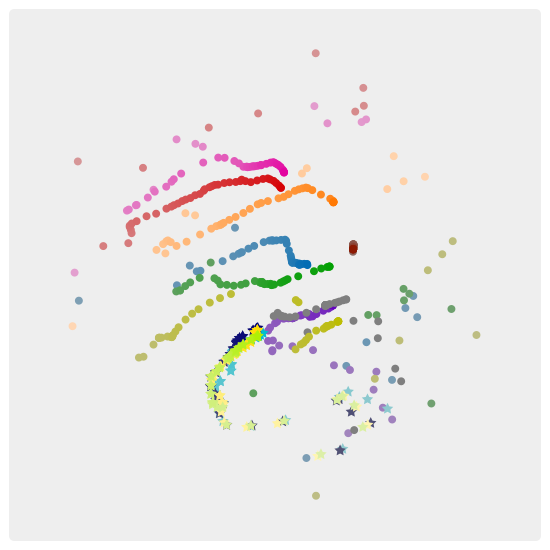}}
    \subfigure[S4 w. subdiv. patches w. exp.\label{subfig:patch-s4smallwexp}]{\includegraphics[width=0.19\textwidth]{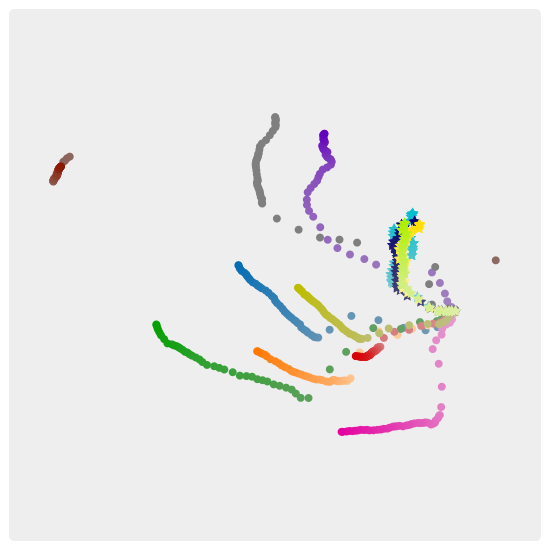}}
    \subfigure[Euclidean w. exp.\label{subfig:patch-edwexp}]{\includegraphics[width=0.19\textwidth]{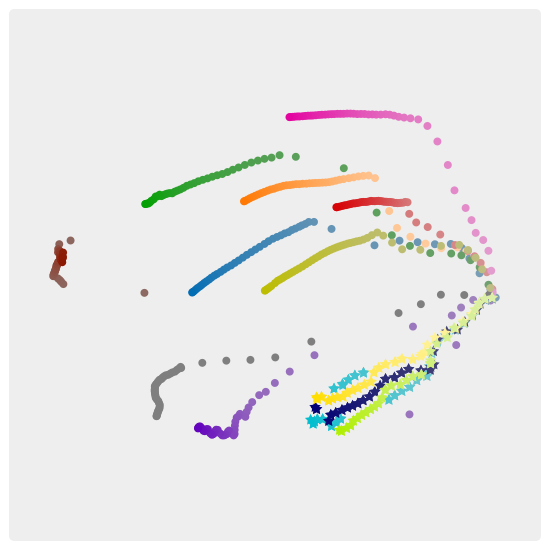}}
     \subfigure[Manhattan w. exp.\label{subfig:patch-manwexp}]{\includegraphics[width=0.19\textwidth]{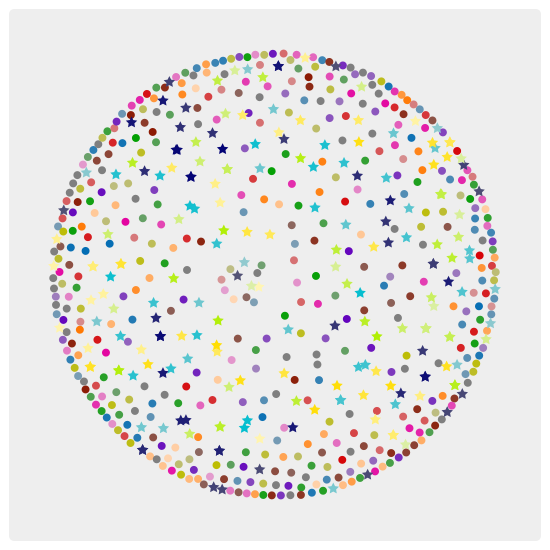}}
    \subfigure[Wasserstein w. exp.\label{subfig:patch-wsswexp}]{\includegraphics[width=0.19\textwidth]{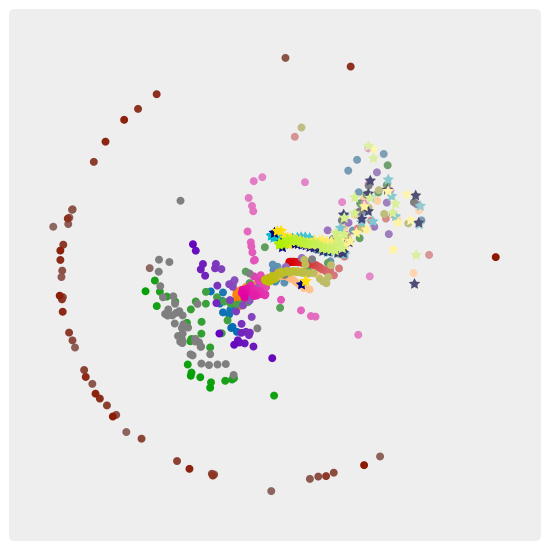}}
    \begin{tabular}{r@{ }l r@{ }l r@{ }l r@{ }l r@{ }l}
        \textcolor{austin}{\textbullet} & austin & \textcolor{csiro}{\textbullet} & csiro & \textcolor{delft-DARSim}{\textbullet} & delft-DARSim & \textcolor{delft-DARTS}{\textbullet} & delft-DARTS & \textcolor{lanl}{\textbullet} & lanl \\
        \textcolor{heriot-watt}{\textbullet} & heriot-watt & \textcolor{melbourne}{\textbullet} & melbourne & \textcolor{stanford}{\textbullet} & stanford & \textcolor{stuttgart}{\textbullet} & stuttgart \\
        \textcolor{experimentrun1}{\ding{72}} & exp. run1 & \textcolor{experimentrun2}{\ding{72}} & exp. run2 & \textcolor{experimentrun3}{\ding{72}} & exp. run3 & \textcolor{experimentrun4}{\ding{72}} & exp. run4
    \end{tabular}
    \caption{
    (a)-(e): 2D plots of the ensemble simulation spatial maps for different similarity metrics with all patches (``patch" mode).
    (f)-(j): same as (a)-(e) but using the segmented spatial maps and including the experimental data in the projection.
    The experimental data is plotted as star shapes instead of circles.
    The segmentation threshold is $0.001$.
    }
    \label{fig:similaritymetricpatchs}
\end{figure*}

Using different similarity metrics helps us to identify several differences in the simulation outcomes. 
\autoref{fig:similaritymetricgroups} shows an overview of the ensemble with and without experimental data in ``group" mode. 
First, we receive a striking observation that the experimental data representation appear close to each other in the overviews with every similarity metric~(Figures~\ref{subfig:s4wexp}-\ref{subfig:wsswexp}). 
This outcome validates the correctness of the overviews. 
Looking closely at the overviews without embedding experimental data, e.g., Figures~\ref{subfig:s4nexp}-\ref{subfig:wssnexp}, we also can see similar pattern deriving from the different metrics, e.g., two sites \textcolor{heriot-watt}{\textit{heriot-watt}} and \textcolor{stanford}{\textit{stanford}} are formed in one group, \textcolor{lanl}{\textit{lanl}} always stands alone, and the other site forms one group with the same spatial arrangement in each view except for Figure~\ref{subfig:wssnexp}. 
We find that these patterns align with the manual visual comparison of the space-time cube renderings.
In~\autoref{subsubsection:similaritfig:similarity_compared_to_spatial_mapsy-view}, we show the projection using the ``S4 with subdivided patches'' side-by-side to the space-time-cube renderings of \austin{austin}, \stuttgart{stuttgart}, \stanford{stanford}, and \heriotwatt{heriot-watt}.
Overall, the projections in~\autoref{fig:similaritymetricgroups} show few differences among the different similarity metrics.
Only the Wasserstein distance shows quite a different distribution of the groups in the projection.

\autoref{fig:similaritymetricpatchs} shows the corresponding projections of the ensemble in ``patch" mode.
For most metrics, the projection of all patches results in prominent time-curves with the time curve arrangements representing a similar pattern to the aggregated counterparts of~\autoref{fig:similaritymetricgroups}.
Though, the time curve representations differ more clearly for different similarity metrics.
For example, the projections for S4 with small patches and Euclidean distance in Figure~\ref{subfig:patch-s4smallnexp} and Figure~\ref{subfig:patch-ednexp}, show a common starting point for all groups.
This is also true when applying segmentation and including the experiment results (Figures~\ref{subfig:patch-s4smallwexp}-\ref{subfig:patch-edwexp}). 
For the projection using Manhattan distance without the experimental data, we also find time curves, but this time with a less clear common starting point (Figure~\ref{subfig:patch-mannexp}).
In contrast to the above, the projection using the S4 metric in Figure~\ref{subfig:patch-s4nexp} differentiates the groups quite well, but it does not reveal any common starting point.
We also find no clear time-curves or starting point for the Manhattan distance with experimental data (Figure~\ref{subfig:patch-manwexp}), as well as in both projections which use the Wasserstein distance (Figures~\ref{subfig:patch-wssnexp} and~\ref{subfig:patch-wsswexp}).

\begin{figure*}
    \subfigure[2D projection of patches via MDS and
S4 with subdiv. patches.\label{subfig:2dmdsS4smallEnhanced}]{\includegraphics[width=0.24\textwidth]{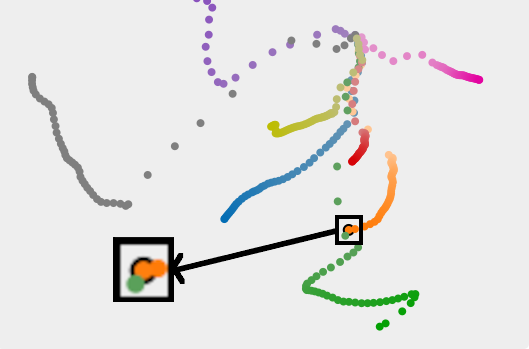}}
    \subfigure[\csiro{Csiro} patch rendering at $24$\,h.]{\includegraphics[width=0.37\textwidth]{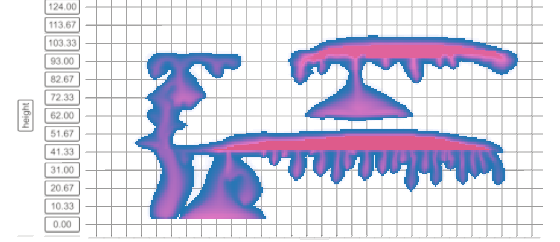}}
    \subfigure[\delftDARSim{Delft-DARSim} patch rendering at $5$\,h.]{\includegraphics[width=0.37\textwidth]{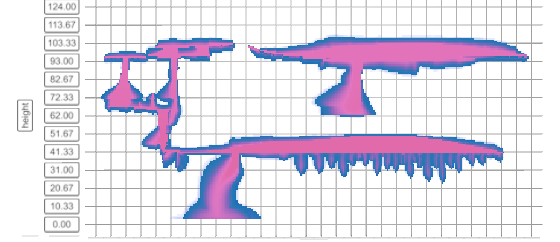}}
    \caption{(a) shows the MDS projection of individual patches with enlarged cutout of last \csiro{csiro} patch at $24$\,h (b) and \delftDARSim{delf-DARSim} patch at $5$\,h.
    Because both patches are from completely different time steps, this suggests that the simulation of \delftDARSim{delft-DARSim} progresses faster than the simulation of \csiro{csiro}.
    }
    \label{fig:csiro-and-darsim-similar-at-different-time-steps-similar-at-different-time-steps}
\end{figure*}

We find that the projections which use the ``S4 with subdivided patches'' similarity metric match closest to the perceived differences in the spatial maps.  
The projections which use Euclidean distance show similar patterns.
The time curve overview in Figures~\ref{subfig:patch-s4smallnexp} and~\ref{subfig:patch-ednexp} reveal that the simulations progresses at different speeds.
The time curves of the simulations all start close to each other and move away from the center over time.
While some take big steps to extent far from the center like \delftDARSim{delft-DARSim} and \stanford{stanford}, others like \csiro{csiro} and \delftDARTS{delft-DARTS} stay rather close.
For example, the projection in Figure~\ref{subfig:patch-s4smallnexp} suggests that the last time step of \csiro{csiro} is one of the closest time steps to \delftDARSim{delft-DARSim}. 
Looking at the space-time cube rendering of both patches side-by-side, we can visually verify that the last patch of \csiro{csiro} at $24$\,h is indeed quite similar to \delftDARSim{delft-DARSim} at $5$\,h (\autoref{fig:csiro-and-darsim-similar-at-different-time-steps-similar-at-different-time-steps}). 
After $5$\,h, \delftDARSim{delft-DARSim}'s simulation progresses and changes further which is also reflected in the simulation. 
Given the matching between the context view applying the two metrics and the detail view, we argue that the two similarity metrics work best with our data.

\subsection{Visual Comparison with Experiment Groups -- \textbf{Q2}}
\begin{figure*}
    \subfigure[Threshold $0$.]{\includegraphics[width=0.26\textwidth]{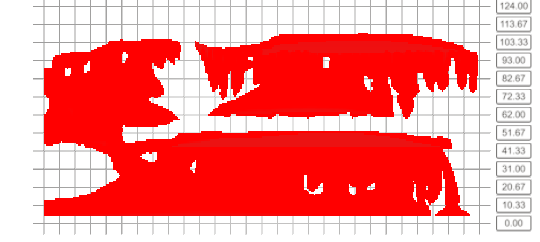}}
    \subfigure[Threshold $0.042$.]{\includegraphics[width=0.26\textwidth]{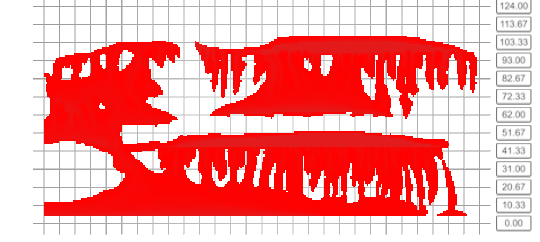}}
    \subfigure[\experimentrunA{experimental run $1$}]{\includegraphics[width=0.26\textwidth]{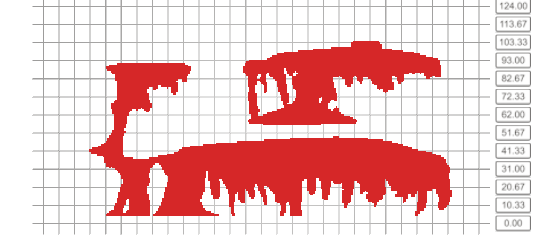}}
    \subfigure[\Tff{}.]{\includegraphics[width=0.19\textwidth]{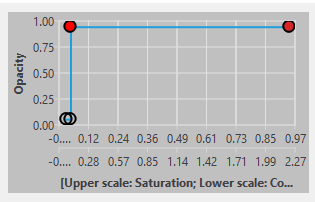}}
    \caption{Side-by-side comparison of the concentration spatial maps of \delftDARSim{delft-DARSim} with a threshold of in (a) $0.0$, and in (b) $0.042$, to the segmentation map of the concentration data of \experimentrunA{experimental run $1$} (c).
    All show a rendering of the last patch at $24$\,h.
    (d): The used \tff{} for selecting the threshold.
    }
    \label{fig:delft-vs-exp1-th}
\end{figure*}

The main difference when comparing the simulation groups with each other versus the comparison including the experimental group is that the experimental data is only available to us as segmentation maps.
In the experimental grid, a chemical ingredient was added which changes its color in contact with \CO{}.
This allows to visually detect the presence of \CO{} during image processing, by applying appropriate thresholds on the amount of color per grid cell.
For our computational comparison, we also have to apply a segmentation on the simulation groups based on a specific threshold.
In Figures~\ref{subfig:s4wexp}-\ref{subfig:wsswexp} and Figures~\ref{subfig:patch-s4wexp}-\ref{subfig:patch-wsswexp}, we use a segmentation threshold of $th>0.001$ and include the experiment runs in the projection.

We find in ``group" mode (Figures~\ref{subfig:s4wexp}-\ref{subfig:wsswexp}) that the experimental runs form one cluster that is distant from the rest of the groups.
When projecting all patches (Figures~\ref{subfig:patch-s4wexp}-\ref{subfig:patch-wsswexp}), the experimental runs still form a separate cluster and stick together.
However, they appear to be rather close to the groups \stanford{stanford} and \heriotwatt{heriot-watt}, though, only for the very first few patches, after which \stanford{stanford} and \heriotwatt{heriot-watt} then quickly diverge from them. 

At this point, we cannot say whether any of the groups are actually close to the experiments based on our metrics and projections.
Our ML-model was not trained on segmentation data and we cannot expect the results to be reliable for such data.
Applying the Wasserstein metric on segmentation maps essentially comes down to counting ones and zeros and comparing the countings among the groups which ignores any potential shapes in the spatial maps.
While Manhattan distance and Euclidean distance are better suited for comparing spatial shapes, they are still heavily affected from the chosen threshold for the segmentation.

Therefore, we instead try to verify the projection by visually comparing the experimental runs with the different simulation groups in the space-time cube renderings.
By adjusting the \tff{} to a function which maps all concentration values to either fully transparent or fully opaque colors depending on a threshold, we can see how the simulation data would look like if it were transformed to segmentation maps by applying this threshold. 
Thus, we can visually figure out which threshold, if applied to the simulation data, most closely resembles the experimental data. 
For example, \delftDARSim{delft-DARSim} visually resembles the \experimentrunA{experiment run 1} at $24$\,h far better with a threshold of $0.042$ on the concentration data instead of a threshold of $0.001$~(\autoref{fig:delft-vs-exp1-th}).
Though, we do not find any suitable threshold for neither \stanford{stanford} nor \heriotwatt{heriot-watt}.
We find that both, especially \stanford{stanford}, seem to be quite different to the experiments and that the other groups, with the exception of \lanl{lanl}, look more similar to the \spatiotemp{} behavior of the experiment runs.

\subsection{Correlation between Saturation, Concentration, and Time Series Data -- \textbf{Q3}}
\begin{figure*}
    \subfigure[MDS, S4 w. subdiv. patches, saturation.\label{subfig:mdss4smallsat}]{\includegraphics[width=0.49\textwidth]{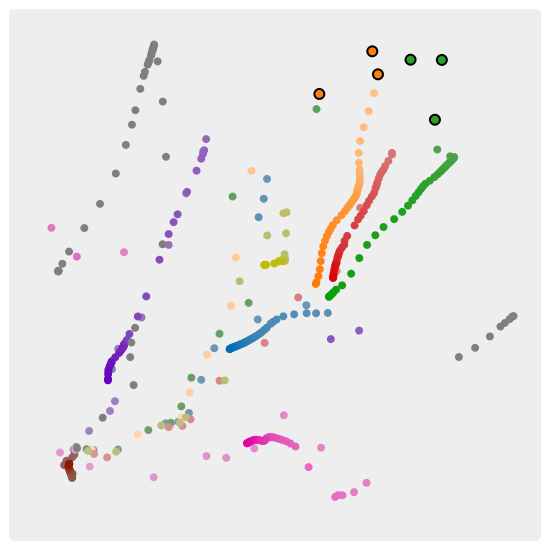}}
    \subfigure[MDS, S4 w. subdiv. patches, concentration.\label{subfig:mdss4smallcon}]{\includegraphics[width=0.49\textwidth]{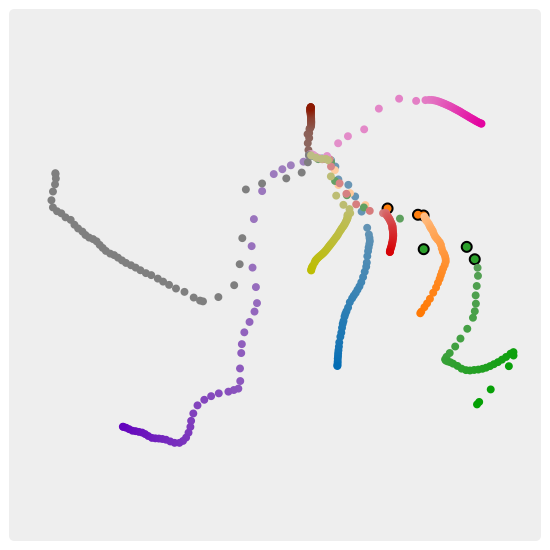}}
    \caption{
    The MDS projection of the individual patches of the simulation ensemble for \csiro{csiro} and \delftDARSim{delft-DARSim}.
    The three patches around the injection stop (after $5$\,h) are highlighted in (a) and (b) with black halos.
    The projection in (a) uses only saturation, in (b) only concentration data.
    After injection stop, the time-curves in (a) make a turn back to the origin, whereas in (b), they still diverge, though slower.
    }
    \label{fig:concentration-vs-saturation}
\end{figure*}
\noindent
\textbf{Temporal Differences between Saturation and Concentration:}
By using saturation and concentration separately in similarity metric computation, we find significant differences in the temporal behavior of saturation compared to concentration.
Figure~\ref{fig:concentration-vs-saturation} shows two MDS projections using the ``S4 with subdivided patches'' as similarity metric.
The projection in Figure~\ref{subfig:mdss4smallsat} with ``patch" mode relates to the saturation data and Figure~\ref{subfig:mdss4smallcon} shows the projection of the concentration data.
We find that in both projections, the time curves have a common point of origin but behave increasingly differently throughout the simulation.
For the concentration data, the time curves keeps mostly moving away from the center, such that the latest patches are one of the out-most points in the projection.
However, for the saturation data, they first briefly diverge, but then make a turn back in the direction of the origin. 
By interacting with the similarity view and the line charts views, we find that this turnaround happens right after injection stop at $t=5$\,h when the saturation, i.e., mobile\,\CO{}, has reached its maximum in all of the boxes. 
Hovering over the saturation maximum data of any of the boxes highlights the most distant patches in the similarity view (Figure~\ref{subfig:mdss4smallsat}).
The mobile\,\CO{} only decreases after injection stop, which explains that the patches after this point become more similar again to the previous patches.
We validate this by visually inspecting the space-time cube renderings for the saturation data.

Although the patches in the concentration plot keep diverging, they diverge much slower after the injection stop, which can be seen in Figure~\ref{subfig:mdss4smallcon} where the patches around the injection stop are highlighted via black halos.

\begin{figure*}
    \subfigure[\boxb{} saturation of \delftDARSim{delft-DARSim}.\label{subfig:boxbsatdelft-darsim}]{\includegraphics[width=0.24\textwidth]{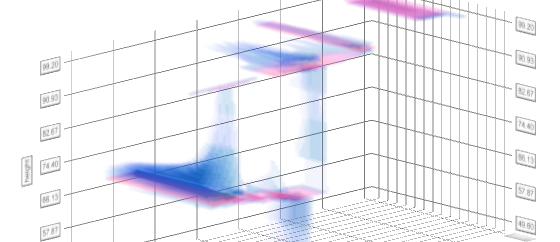}}
    \subfigure[\boxb{} saturation of \melbourne{melbourne}.]{\includegraphics[width=0.24\textwidth]{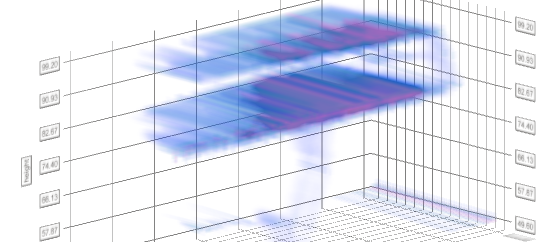}}
    \subfigure[\boxb{} con. of \delftDARSim{delft-DARSim}.\label{subfig:boxbcondelft-darsim}]{\includegraphics[width=0.24\textwidth]{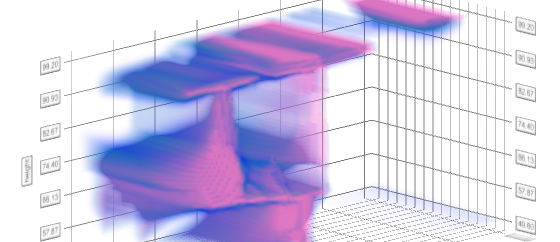}}
    \subfigure[\boxb{} concentration of \melbourne{melbourne}.]{\includegraphics[width=0.24\textwidth]{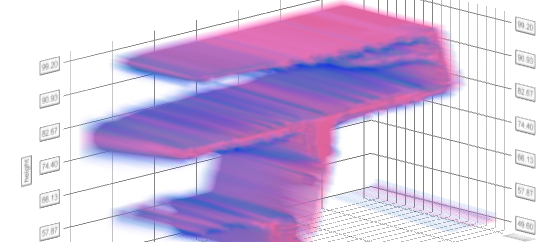}}
    \subfigure[Time series of mobile \CO{} in \boxb{} for \delftDARSim{delft-DARSim} and \melbourne{melbourne}.]{\includegraphics[width=0.49\textwidth]{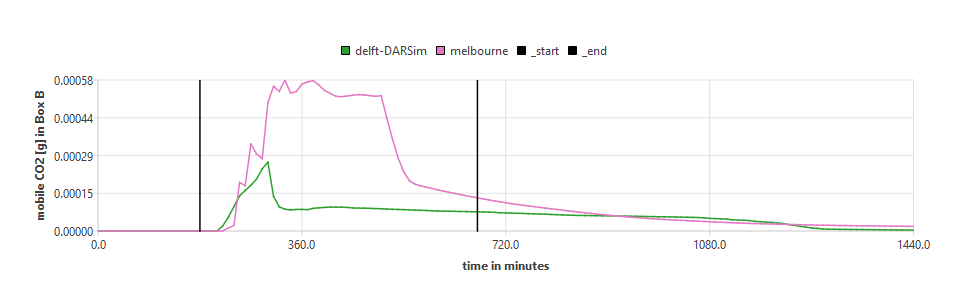}}
    \subfigure[Time series of dissolved \CO{} in \boxb{} for \delftDARSim{delft-DARSim} and \melbourne{melbourne}.]{\includegraphics[width=0.49\textwidth]{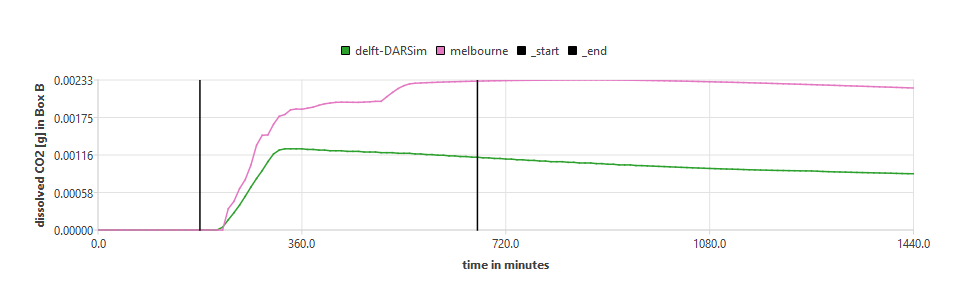}}
    \caption{
    The space-time cube visualizations of \boxb{} for \delftDARSim{delft-DARSim} and \melbourne{melbourne} (a)-(d) match the trend of the corresponding mobile\,\CO{} and dissolved\,\CO{} time series (e)-(f).
    Subfigures (a) and (b) show the suddenly increasing saturation in \boxb{} after reaching the spilling point as well as the decreasing saturation intensity over time as more and more \CO{} dissolves, thus, concentration increases (see (c) and (d)).
    }
    \label{fig:melb-darsim-nice}
\end{figure*}
\noindent
\textbf{Correlation between Spatial Maps and Time Series Data:}
By zooming to the boxes of interest and interacting with the time series views and the space-time cube renderings, we can inspect the boxes of interest in more detail and detect when certain events happen.
For example, the visual inspection of \boxb{} allows to see when the spilling point of \boxa{} is reached.
When \boxa{} has reached its maximum capacity of \CO{} gas, injecting more \CO{} results the \CO{} gas to ``spill'' over the spilling point and leak into \boxb{} through a coarse-grained riff ~(\autoref{fig:benchmark_geometry} to the left of \boxa{}).
If this happens, we can see increasing \CO{} saturation and concentration in the space-time cube rendering of \boxb{} which should also correlate to increasing mobile\,\CO{} and dissolved\,\CO{} time series data in \boxb{}.
We find that this correlation between spatial maps and time series data matches well, which we show for groups \delftDARSim{delft-DARSim} and \melbourne{melbourne} in~\autoref{fig:melb-darsim-nice}.
There, we notice that the first saturation in \boxb{} is visible after $220$\,minutes for \delftDARSim{delft-DARSim} and after $230$\,minutes for \melbourne{melbourne} (not considering outlier \lanl{lanl}).
Most other groups start showing \CO{} saturation at around $250$\,minutes in \boxb{}.

While it would be possible to use the time series data itself for detecting when the spilling point is reached, the visualization provides an convenient approach to validate the time series data and that the measured \CO{} in \boxb{} is indeed due to leakage from \boxa{}.
Furthermore, this approach also provides means to quickly identify when the spilling point is reached for the experiment runs for which no time series data is available to us.
We identify the time of reaching the spilling point to be after $250$\,min, $260$\,min, $270$\,min, $270$\,min for the experiment runs 5, 2, 1, and 3 respectively.

\subsection{Shape and Development of Fingers Differ Throughout the Ensemble -- \textbf{Q4}}\label{subsection:dev-and-shape-of-fingers}
\begin{table}[ht]
    \centering
    \setlength\tabcolsep{1.5pt} 
    \begin{tabular}[t]{c}
        \multicolumn{1}{c}{Groups} \\ \hline
         \\
        \hline
         \stuttgart{Stuttgart}\\ \hline
         \stanford{Stanford} \\ \hline
         \melbourne{Melbourne} \\ \hline
         \heriotwatt{Heriot-Watt} \\ \hline
         \delftDARSim{Delft-DARSim} \\ \hline
         \delftDARTS{Delft-DARTS} \\ \hline
         \csiro{Csiro} \\ \hline
         \austin{Austin} \\ \hline
         \experimentrunA{Experiment Run 1} \\ 
    \end{tabular}
    \begin{tabular}[t]{c|c|c}
        \multicolumn{3}{c}{Shape} \\ \hline
        Thin & Wide & Diffusive\\
        \hline
        \stuttgart{\textbullet} & & \\ \hline
        & & \stanford{\textbullet} \\ \hline
        & \melbourne{\textbullet} & \\ \hline
        \heriotwatt{\textbullet} & & \\ \hline
        \delftDARSim{\textbullet} & & \\ \hline
        \delftDARTS{\textbullet} & & \\ \hline
        \csiro{\textbullet} & & \\ \hline
        \austin{\textbullet} & & \\ \hline
        \experimentrunA{\textbullet} & & \\ 
    \end{tabular}
    \begin{tabular}[t]{c|c|c}
        \multicolumn{2}{c}{Length} \\ \hline
        Short & Long \\
        \hline
        & \stuttgart{\textbullet}  \\ \hline
        \stanford{\textbullet} & \\ \hline
        \melbourne{\textbullet} & \\ \hline
        & \heriotwatt{\textbullet} \\ \hline
        & \delftDARSim{\textbullet} \\ \hline
        & \delftDARTS{\textbullet} \\ \hline
        & \csiro{\textbullet} \\ \hline
        & \austin{\textbullet} \\ \hline
        & \experimentrunA{\textbullet} \\ 
    \end{tabular}
    \begin{tabular}[t]{c|c|c}
        \multicolumn{3}{c}{Development Behavior} \\ \hline
        Initial & Recurring & Continuous\\
        \hline
        & \stuttgart{\textbullet} & \\ \hline
        \stanford{\textbullet} & & \\ \hline
        & & \melbourne{\textbullet} \\ \hline
        & \heriotwatt{\textbullet} & \\ \hline
        & \delftDARSim{\textbullet} & \\ \hline
        & & \delftDARTS{\textbullet} \\ \hline
        & \csiro{\textbullet} & \\ \hline
        & & \austin{\textbullet} \\ \hline
        & & \experimentrunA{\textbullet} \\ 
    \end{tabular}
    \caption{Classification of the finger development across the groups by overall shape, length, and behavior.
    We classify the experiment runs the same and list \experimentrunA{experiment run 1} as representative for all experiment runs in the table.
    \label{tab:results-finger-shape-size-behavior}
    }
\end{table}
\begin{figure*}
    \subfigure[\stanford{Stanford} fast and diffusive finger development.]{\includegraphics[width=0.36\textwidth]{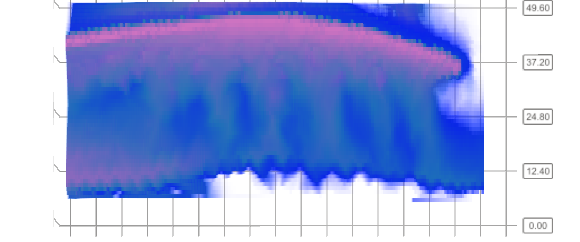}}
    \subfigure[\heriotwatt{Heriot-watt} thin and long but pulsing fingers.]{\includegraphics[width=0.36\textwidth]{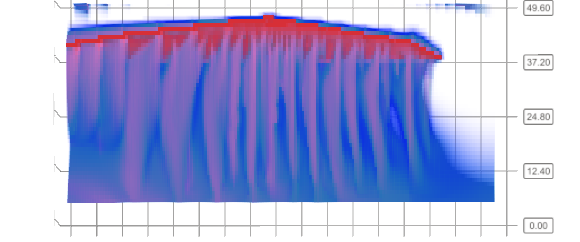}}
    \subfigure[\heriotwatt{Heriot-watt} saturation.\label{subfig:heriot-watt-saturation}]{\includegraphics[width=0.21\textwidth]{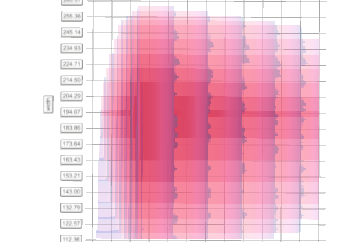}}
    \subfigure[\stuttgart{Stuttgart} thin and slow finger development.]{\includegraphics[width=0.36\textwidth]{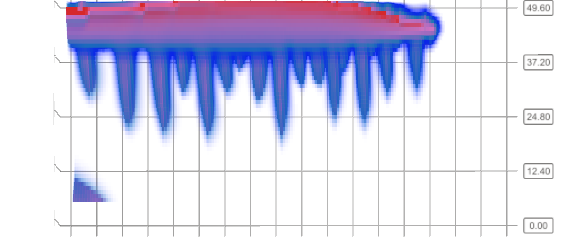}}
    \subfigure[\melbourne{Melbourne} slow and rather wide/round finger development.]{\includegraphics[width=0.36\textwidth]{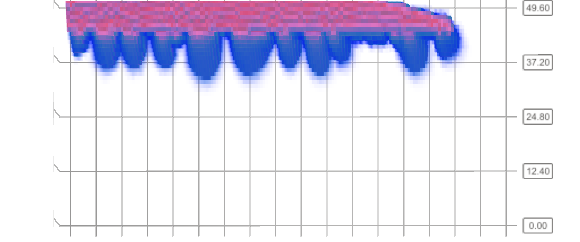}}
    \subfigure[\melbourne{Melbourne} saturation.\label{subfig:melbourne-continuous}]{\includegraphics[width=0.21\textwidth]{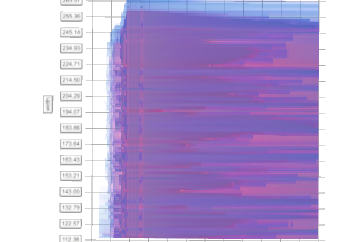}}
    \caption{
    The volume visualization of \boxa{} for the first $24$\,h reveals different types of finger development.
    For example, some groups show a ``pulsing'' finger development.
    We can also see this easily by slicing into the volume to \boxa{} and adjusting the \tff{} accordingly (c).
    }
    \label{fig:pulsing-fingers}
\end{figure*}

By slicing into the volume for \boxa{}, we find different shapes and types of developments of the fingers that are visible by looking at the concentration data of the different groups.
We categorize them by overall shape, length, and development behavior.
The individual categories are further subdivided as follows: overall shape in ``thin'', ``wide'', and ``diffusive''; length in ``short'' or ``long''; and development behavior in ``initial pulse'', ``recurring pulses'', and ``continuous pulses''.
We provide our classification in~\autoref{tab:results-finger-shape-size-behavior}.
Based on our classification, we find that most groups develop thin and long fingers, and consider only \stanford{stanford} and \melbourne{melbourne} to have neither long nor thin fingers.
\melbourne{Melbourne} has rather short and wide fingers, while for \stanford{stanford} they appear short and diffusive.
With ``diffusive'' we describe the fingers to have no clear shape in the early stages, and that the distribution of \CO{} concentration in \boxa{} seems to be fuzzy.

During the analysis of the fingers in \boxa{}, we noticed that instead of the \CO{} gradually dissolving into the water and dropping down in the form of fingers, this dissolving process often happens in the form of ``pulses''.
We therefore classify the groups in ``initial'', ``recurring'' and ``continuous'' regarding the development behavior of fingers.
With ``initial'' we refer to groups for which the fingers develop suddenly and only once in one initial ``drop''.
``Recurring'' refers to groups for which we noticed recurring pulses, where each pulse introduces more \CO{} concentration which developed in a short time period.
``Continuous'' refers to the groups where we cannot find recurring behavior or an initial fast dop, but for which the \CO{} instead gradually dissolves into fingers which thus continuously grow.

We find that only \stanford{stanford} has one initial fast drop.
The other groups have either recurring pulses or a continuous growth of the fingers.

We provide some examples for different fingers in~\autoref{fig:pulsing-fingers}.
After further investigating the pulsing behavior, we also find that we see a brief period of low saturation during each pulse for the groups which have pulsing behavior, as we show in Figure~\ref{subfig:heriot-watt-saturation}.
Compared to groups with continuous finger development, we see a continuous trail of low saturation over time instead, as we show in Figure~\ref{subfig:melbourne-continuous}. \ms{nice results section!}

\section{Discussion}
In this section we discuss some strengths and shortcomings and potential future work.
\subsection{Strengths and Shortcomings}

\noindent
\textbf{Similarity Metric: }
In the previous section, we provided multiple results, which show the utility of our visual approach.
Our visual approach incorporates multiple commonly used similarity metrics like Euclidean, Manhattan, and Wasserstein distances.
One strength of our approach is the incorporation of multiple such similarity metrics, but also of a more sophisticated ML approach as similarity metric, which can be used to project the full ensemble into an overview.
While the incorporation of the ML model as a similarity metric allows integrating a metric that should consider features in the data such as \spatiotemp{} behavior, it is not clear what the model actually computes, as it acts as a black box to us.
Furthermore, it may not be suited to be used on segmentation data, although it produces promising results. 

\noindent
\textbf{Dimensionality Reduction: }
Another strength is the interactive linking between our overview, space-time cube renderings, and line charts.
While the overview in the projection view allows to quickly understand similar groups overall and how the simulations of the groups temporally relate to each other, it hides all the details of the actual simulation data.
Also, the overview does not show the projection quality such as the remaining stress of the MDS projection.
We addressed both issues by linking the views in our visual approach which enables efficient navigation and exploration of the ensemble dataset on different levels of detail by brushing in the overview or line charts.
Hence, it is possible to manually verify the correctness of the projections to a certain degree by visually comparing the data details.
Though, a proper visualization of the projection quality might be a good fit for future work.

\noindent
\textbf{Volume Rendering: }
One additional benefit in our visual approach was the selection of space-time cubes for the visual representation of the spatial maps.
A space-time cube is well suited for 2D+T data to provide a static overview of multiple 2D spatial maps that shows how saturation and concentration propagates through the geometry.
However, a \tff{} is necessary to properly leverage the capabilities of a space-time cube representation and configuring a \tff{} can be difficult.
It introduces many pitfalls in the visual analysis when a \tff{} is not properly configured.
Nevertheless, we showed the flexibility of the space-time cube rendering with an interactive \tff{} and how it can be used to visually compare simulation to experiment runs with different thresholds on the saturation or concentration data.

\noindent
\textbf{Generalization: }
Even though we developed our approach in regard to the data of the benchmark study, it can be generalized to other data as well.
Proper pre-processing as described in~\autoref{subsubsection:data-preprocessing} might be required.
For the overview, it is just a matter of choosing a similarity metric that is capable of computing the similarity between patches in the data that relate to a time component.
The concept of patches can easily be extended to 3D+T(ime)+V(ariables)  data which can also be processed by our chosen ML model.
A drawback is that the ML model has to be trained on the data before it can be used with our visual approach and that too little data and poor hyperparameter settings can introduce overfitting.
Euclidean, Manhattan, and Wasserstein distances do not have these issues and are still viable options for 3D+T+V data.

One core strength of choosing a space-time cube for visualizing the data is that we can show a static overview of multiple time steps at once, even for 2D+T data.
We can still use the same concept for 3D+T data, although at least one dimension has to be collapsed or limited to one slice of it before we can render the remaining data as a 3D volume.
Furthermore, our visual approach currently visualizes only one variables of the simulation data per space-time cube rendering.
In our implementation, we chose two such renderings as juxtaposed views, which allows a side-by-side comparison of two space-time cube renderings.
This limits the capabilities of visualizing more than two ensemble members or variables at once.
The interaction with space-time cubes allows to zoom to specific regions of interests that are defined by the benchmark study description.
Other data might not have these specific regions of interests and this interaction should be changed to allow to zoom to arbitrary regions of interests.

\noindent
\textbf{Scalability: }
In terms of ensemble size, we figure that bigger ensembles might introduce visual clutter in the overview and line chart views.
Besides that, our approach lacks techniques for parameter space analysis~\cite{Sedlmair2014VisualParamSpaceAnalysis}, which is a common task for big ensembles.
We discuss options to address these limitations and other future work below.

\subsection{Future Work}\label{subsection:futurework}
The previous discussion and mentioned issues and drawbacks provide multiple directions to extend our visual approach in future work.
For example, it is currently necessary to train the ML model we use as similarity metric.
This model could be replaced with another model that does not require retraining but still captures \spatiotemp{} features, such as~\cite{Huesmann2022SimilarityNet}.

Furthermore, our visual approach should be able to naturally process arbitrary 3D+T+V \spatiotemp{} data.
For a better integration of more variables in the space-time cube visualizations, we propose to superimpose multiple space-time cubes of different variables.
Combined with a \tff{} that can be configured for each space-time cube individually, this superimposition can extend the space-time cube visualization to show multivariate data in a single view.

In addition to our research questions~(\autoref{sub:tasks}), the domain experts have expressed great interest to not only discover and analyze the outcome but also understand the reason behind why simulation outcomes vary.
However, the differences between two model runs can be manifold and range from possibly different underlying balance equations, discretization approaches and constitutive relations over varying spatial parameters and grid resolutions up to diverse numerical solution approaches and parameters.
Therefore, we leave this research question for future work where possibly more selected variations of a computational model are evaluated. 

Our current work focused on an in-depth analysis of the ensemble of different simulation runs.
Our hope is that this analysis will serve as a starting point for the next steps to agglomerate the ensemble into joint insights and decisions.
For that, further extensions of our approach to make it amenable for broad communication will be necessary~\cite{munzner2014visualization}.
Meaningful physical interpretations of the applied similarity metrics will be of particular interest.

\section{Conclusion}
We presented an approach for the interactive visual analysis of simulation codes and experiment ensembles of porous media fluid flows.
An overview with a variety of similarity metrics allows to identify and compare \spatiotemp{} patterns, such as 
the differences in the development of gravity fingers, 
and also to identify correlations among attributes measured from the spatial maps, such as the mobile\,\CO{} and dissolved\,\CO{} in different regions of interest.
Detail views display \CO{} concentration and saturation in a space-time cube format, and support the navigation through the ensemble data. 

We applied our approach to data from a benchmark study with nine different simulation models. 
Our analysis revealed new insights into 
ranking of simulation results with respect to experimental data,
correlation between \CO{} saturation and concentration,
and gravity finger development.
As next steps, we plan to expand our collaboration to involve domain experts 
from all nine research sites and to jointly derive decisions and lessons learned from this large scale simulation endeavor. 

\backmatter

\bmhead{Acknowledgments}

Funded by Deutsche Forschungsgemeinschaft (DFG, German Research Foundation) under Germany's Excellence Strategy - EXC 2075 – 390740016 and Project Number 327154368 – SFB 1313.
We acknowledge the support by the Stuttgart Center for Simulation Science (SimTech).  Parts of this work have been done in the context of CEDAS, the Center for Data Science at the University of Bergen.

\section*{Declarations}

\subsection*{Funding}
Funded by Deutsche Forschungsgemeinschaft (DFG, German Research Foundation) under Germany’s Excellence Strategy - EXC2075 – 390740016 and Project Number 327154368 – SFB 1313.

\subsection*{Competing interest}
The authors have no relevant financial or non-financial interests to disclose.

\subsection*{Author contributions}
All authors contributed to the conceptual design of our work, and made substantial contributions in writing and revising this manuscript.
The implementation of the concepts, data preparation, analysis, and writing of the first draft were performed by R. Bauer.
All authors read and approved the final manuscript.

\subsubsection*{Corresponding author}
Correspondence to \href{mailto:ruben.bauer@visus.uni-stuttgart.de}{Ruben Bauer}.

\subsection*{Data availability}
The datasets which are used in this manuscript, mainly the benchmark study results, are available in the Fluidflower repositories, \url{https://github.com/fluidflower}.

\bibliography{sn-bibliography}

\end{document}